\documentclass[aps,twocolumn,preprintnumbers,nofootinbib]{revtex4-1}

\usepackage{subcaption}
\usepackage{cancel}
\usepackage[utf8]{inputenc}
\usepackage[dvipdf]{graphicx}
\usepackage{xcolor}
\usepackage{color}
\usepackage{relsize}
\usepackage{graphics}
\usepackage{epstopdf}
\usepackage{hyperref}
\usepackage{mathrsfs}
\usepackage{amssymb}
\usepackage{amsthm}
\usepackage{amsmath}
\usepackage{hyperref}

\begin{document}

%
%

\preprint{}
\preprint{}

\vspace*{1mm}

\title{The Inflaton Portal to a Highly Decoupled EeV Dark Matter Particle }

\author{Lucien Heurtier$^{a}$}
\email{heurtier@email.arizona.edu}
\author{Fei Huang$^{a}$}
\email{huangfei@email.arizona.edu}

\vspace{0.1cm}
\affiliation{
${}^a$ Department of Physics, University of Arizona, Tucson, AZ   85721}

\begin{abstract} 
We explore the possibility that the dark-matter relic abundance is generated in a context where the inflaton is the only mediator between the visible and the hidden sectors of our universe. Due to the relatively large mass of the inflaton field, such a portal leads to an extremely feeble interaction between the dark and the visible sectors suggesting that the dark sector cannot reach any thermal equilibrium with the visible sector. After the two sectors are populated by the decay of the inflaton, a heavy dark-matter particle thermally decouples within the dark sector. Later, a lighter dark particle, whose decay width is naturally suppressed by the inflaton propagator, decays into the visible sector after it dominates the energy density of universe. This process dilutes the dark-matter relic density by injecting entropy in the visible sector. We show that an inflaton mass of $\mathcal{O}(10^{13})$ GeV together with couplings of order one are fully compatible with a dark-matter relic abundance $\Omega h^2\sim 0.1$. As a general feature of the model, the entropy dilution mechanism is accompanied by a period of early matter domination, which modifies the amount of e-folds of inflation necessary to accommodate Planck data. Moreover, the coupling of the inflaton to the dark and visible sectors brings loop contributions to the inflationary potential which can destabilize the inflation trajectory. Considering all these complementary constraints, we show that, in the context of a plateau-inflation scenario such as the $\alpha$-attractor model, the inflaton can constitute a viable mediator between the visible sector and a $\sim 10$ EeV dark-matter candidate. Furthermore, we show that improved constraints on the tensor-to-scalar ratio and spectral index could potentially rule out dark-matter scenarios of this sort in the future.
\end{abstract}

\maketitle


\maketitle

\setcounter{equation}{0}



\section{Introduction}

\noindent
Although the Standard Model (SM) appears to be one of the most complete and accurate theory of particle physics in the last decades, a few major caveats remain to be addressed, especially when particle physics is discussed in the context of cosmology. On the one hand, the rotation curves of galaxies \cite{Rubin:1970zza}, the observation of the Bullet Cluster \cite{Clowe:2003tk}, and the study of the Cosmic Microwave Background (CMB) \cite{Ade:2015xua, Hinshaw:2012aka} suggest that our universe contains a significant fraction of dark matter (DM). On the other hand, the incredible homogeneity and flatness of our observable universe revealed by analysis of the CMB spectrum renders the vanilla Big Bang Theory somehow obsolete at early time and suggests that the universe underwent a rapid phase of expansion called {\em cosmic inflation}. The problem of dark matter requires the existence of a particle stable on scales longer than the age of the Universe. Moreover a scalar field slow-rolling in a flat enough potential (called the {\em inflaton}) at primordial stages of the universe evolution can produce the desired expansion for diluting inhomogeneities and residual curvature.

During the inflationary phase our spacetime is effectively de Sitter. When inflation ends the inflaton oscillates in a cold and matter dominated universe. The transition between this post-inflationary phase and the thermal history of the universe succeeding it -- referred to as {\em reheating} --  is understood as an out-of-equilibrium decay of the inflaton field, converting its potential energy into a relativistic thermal bath.  The temperature at which the reheating happens is almost unconstrained by theory, the only requirement being that the universe is not reheated below the Big Bang Nucleosynthesis (BBN) temperature ($T_\text{RH}\gtrsim 10$ MeV) and by definition it cannot exceed the energy scale of inflation \cite{Domcke:2015iaa}. In a supersymmetric context gravitino production  at early time may overclose the universe, imposing an upper bound on the reheating temperature usually around $T_\text{RH}\lesssim 10^{(10-12)}$ GeV (see, e.g. \cite{Benakli:2017whb, Dudas:2017rpa, Moroi:1993mb}).

Most of the time the discussion of dark-matter production is disconnected from any explicit formulation of the reheating transition. In the context of {the thermal {\em freeze-out} scenario} however, such an approach is not problematic. Indeed dark matter is in this case in thermal equilibrium with the visible sector before it dynamically decouples, and thermalization of the inflaton decay products erases any specificity concerning the way the reheating takes place. However, in alternative scenarios where dark matter may be produced {non-thermally or out of equilibrium} -- such as the so called {\em freeze-in} scenario -- the way the inflaton decays preferentially into dark matter or into visible particles may have dramatic consequences on the upcoming dark matter production process \cite{Hall:2009bx, Chu:2011be, Chu:2013jja, Dev:2014tla, Bernal:2017kxu, Bhattacharyya:2018evo}. {In particular, in the freeze-in scenario the coupling of the inflaton to SM states which thereafter produce DM particles, necessarily provides a direct decay channel for the inflaton into DM particles at the loop level. It was shown in Ref.~\cite{Kaneta:2019zgw} that such direct decay, which is by construction present in many models of gravitino DM, might contribute significantly to the overall DM abundance, and possibly overclose the universe in certain situations.} Amongst the thermal scenarios, whereas a large class of model remains phenomenologically viable (see, e.g. Refs~\cite{Feng:2008ya,Dienes:2017zjq}), the popular WIMP paradigm for dark-matter production is more and more constrained by direct detection experiments \cite{Cheung:2012gi, Das:2011yr, Shan:2011ct, Pato:2010zk, Bertone:2010rv, Bernal:2008zk, Mena:2007ty, PalomaresRuiz:2010pn, Konar:2009ae}. Therefore, investigating the possible consequences of an explicit reheating model on alternative ways of producing dark matter is becoming particularly relevant. 
{ 
In this paper we study the possibility that the inflaton field --- which is implicitly present in most of the models of Beyond-the-Standard Model (BSM)  cosmology --- is the only mediator between the visible and the hidden sectors. Since the mass of the inflaton is predicted by a large class of models to be of order $10^{13}$ GeV or larger, such interactions which take place through the inflaton portal is expected to be extremely suppressed, suggesting that the dark sector is highly decoupled from the visible sector at very early time.}

Indeed, in Refs.~\cite{dev} and \cite{moi} it was shown that introducing the inflaton as a mediator in the case of thermal and non-thermal scenarios of dark-matter production is not fully satisfactory. In the case of thermal production, the annihilation cross section would be far too suppressed, resulting in an overclosure of the universe. In the case of non-thermal production, $\mathcal O(1)$ couplings  between the inflaton and the dark sector lead to a situation in which the dominant contribution to the dark-matter abundance occurs during reheating and not through the freeze-in mechanism, rendering such a scenario highly fine-tuned. An attempt was proposed in Ref.~\cite{moi} to motivate the hierarchy of couplings necessary to make a freeze-in scenario viable. However, even in a scenario of this sort, obtaining an appropriate annihilation cross section requires the use of a very small parameter, rendering the scenario as unnatural as many of the usual dark matter constructions.

{In Refs.~\cite{hooper1, hooper2} it has been shown that a dark-matter candidate which decouples thermally within a highly decoupled dark sector can dynamically be produced if an appropriate  entropy dilution mechanism re-adjusts its relic abundance. Such mechanism generically requires a late, out-of-equilibrium decay of some dark-sector particle(s) into SM particles. This type of scenario was proposed in Refs. \cite{hooper1, hooper2} in which the dark-matter mass is required to be as large as a few PeV. However, the late-time decay required by the entropy dilution process is  achieved by a significant fine-tuning of the parameters. Moreover the amount of energy density contained in the dark and visible sectors once inflation ends are arbitrarily chosen.}

{
For these two reasons, the inflaton portal turns out to be a perfect candidate to solve in one stroke these two issues. We will indeed show that using the inflaton as a mediator between the hidden sector and the SM bath will get rid of any arbitrary choice of initial conditions and naturally relate the inflationary sector to the physics of dark-matter production. Specifying the full Lagrangian including {the inflationary sector, the dark sector, and the coupling of the inflaton to both dark- and visible-sector particles}, provides a complete set of parameters to describe $(i)$ the decay of the inflaton into the dark sector and the visible sector, $(ii)$ the entropy dilution factor, and therefore the relic density of dark matter, and $(iii)$ the lifetime of the lightest hidden-sector particle, meaning the temperature at which the visible bath will be reheated after its decay. We will furthermore derive useful analytic formulae for the dark-matter relic density {and examine the extent to which our model can be {constrained by primordial cosmological considerations.}} {The connection of the inflationary sector to low energy phenomenology \cite{Argurio:2017joe, Rehman:2018nsn, Rehman:2018gnr, Maity:2018dgy, Maity:2018exj, Aravind:2015xst, Dudas:2017kfz, Choubey:2017hsq, Addazi:2017kbx, Daido:2017wwb, Tenkanen:2016jic, Tenkanen:2016idg, Tenkanen:2016twd, Almeida:2018oid, Ballesteros:2016xej, Borah:2018rca} has been shown to be a powerful way to constrain many facets of particle physics through different angles at once. Nevertheless,} it is the first time that the production of a dark-matter particle interacting with the SM exclusively through the exchange of the inflaton is demonstrated to be successful with a natural choice of couplings, and, last but not least, while introducing only one single mass scale in the inflationary sector.

The paper is organized as follows: In Sec.~\ref{sec:TheModel}, we present the model on which we shall focus throughout this work. In Sec.~\ref{sec:Reheating}, we detail how specifying the couplings of the inflaton to the visible and dark sectors provides a way to get rid of any arbitrary choice on the initial conditions for the hidden- and visible-sector temperatures. In Sec.~\ref{sec:RelicDensity}, we compute the relic density of dark matter in this scenario and establish an interesting correspondence between {the value of the dark-matter relic abundance and the inflaton mass, given a natural choice of the model parameters.} In Sec.~\ref{sec:Inflation},  we study the radiative corrections to the inflationary trajectory induced by large coupling of the inflaton to the matter sector. {In Sec.~\ref{sec:EMD}, we consider different inflation scenarios and compute the number of e-folds of inflation, taking into account the effect of an early matter-dominated era in our cosmological scenario. In Sec.~\ref{sec:detection} we discuss the possible detection signals that our model might give rise to.} We conclude in Sec.~\ref{sec:Conclusion} by commenting on the generality of our findings. We also discuss how our minimal model might be extended both in terms of modifications of the inflationary sector of the theory and in terms of how this sector is coupled to the fields of visible sector.
 


\section{The model}
\label{sec:TheModel}
\noindent
In this section we present the model which will be studied throughout this paper. Following Ref.~\cite{hooper2}, we take the dark sector content of the model to consist of a dark-matter fermion $\chi$ in thermal equilibrium at high energies with a dark scalar $S$. The temperature of the dark sector is denoted by $T_h$. Since interaction between the dark sector and the visible bath being extremely feeble, no thermal equilibrium can be reached between the two baths.  The ratio of their temperatures $\xi(T) \equiv T_h/T$, where $T$ is the temperature of the SM bath, depends essentially only on the number densities of the particle species in each sector produced during reheating and on the effective number of degrees of freedom of each such species. 

In our model, in the same spirit as Ref.~\cite{moi}, the only contact between the visible and the dark sector will be through the exchange of an inflaton particle. {We will fix the matter content of our model to be composed of}
\begin{itemize}
\item A dark-matter candidate, taken to be a Dirac fermion, denoted by $\chi$;
\item A dark scalar $S$, which is in equilibrium with dark matter at high energy before dark matter decouples, and which dominates the energy density before it decays to the SM;
\item The inflaton scalar particle $\phi$, which is singlet under the SM gauge groups, and which interacts both with dark matter and with the SM;
\item {A Majorana fermion \footnote{Note that the choice of a Majorana fermion versus a Dirac fermion has no effect on the upcoming discussion.}  $N$ ({\em e.g.} a right-handed neutrino), which is assumed to be the only particle in the visible sector that couples directly to the inflaton. We also assume that it instantaneously thermalizes {with} the SM thermal bath right after being produced from the inflaton decay.}
\end{itemize}
{The total Lagrangian can be written as
\begin{equation} 
\mathcal{L}=\mathcal{L}_\text{inf}+\mathcal{L}_{h}+\mathcal{L}_{v}+\mathcal{L}_{portal}\,,
\end{equation}
where the first three terms describe the different sectors of the model that are, respectively, the inflationary, the hidden and the visible sectors. The last term contains the ``portal" interactions between the inflaton with the two other sectors.  The inflationary sector is assumed to contain a single scalar field $\phi$ whose dynamics is given by the shape of the inflationary potential $V_\text{inf}$ 
\begin{equation}\label{eq:lagrangianInf}
\mathcal{L}_\text{inf}=\frac{(\partial\phi)^2}{2}-V_\text{inf}(\phi)\,.
\end{equation}}
{The choice of a given inflation potential $V_\text{inf}$ will lead to different signatures in the CMB spectrum which we will discuss in Sec.~\ref{sec:detection}. After inflation ends, and the reheating occurred, the inflaton scalar field is stabilized at its minimum of potential, and its mass is given by $m_{\phi}^2~\equiv~V_\text{inf}''|_{\phi=0}$. 
Note that we assume here that the minimum corresponds to a vanishing field value $\phi=0$ for simplicity. In the cases where the inflaton would take a non-zero field value at the minimum, it would source mass terms for the dark-sector and visible-sector particles which the inflaton is coupling to. Therefore, the contribution from the vacuum expectation value of the inflaton would have to be added to the bare masses of the relevant particles. In the most minimal large-field inflation scenarios (such as e.g. chaotic inflation \cite{Linde:1983gd}, Starobinsky or $\alpha$-attractor models \cite{Starobinsky:1983zz,Kofman:1985aw,Whitt:1984pd} , new inflation \cite{Kallosh:2010ug,Kallosh:2007wm,Linde:2011nh}, or natural inflation \cite{Freese:1990rb,Savage:2006tr}), the inflaton mass in the vacuum is constrained by the CMB power spectrum normalization to be of order $10^{13}~\mathrm{GeV}$. Before studying in full details the inflation trajectory later on, we will in what follows fix the inflaton mass to the benchmark value $m_{\phi}=10^{13}\ \mathrm{GeV}$.} While most models of inflation connected to low energy physics {take advantage of the fact that the mass of the inflaton in the vacuum can be different from the second derivative of the potential during inflation}  to {accommodate} experimental constraints, we aim for minimality and introduce only one mass scale to characterize the inflationary dynamics. The effects of modifying this minimal approach via the introduction of other scales, etc., will be discussed in Sect.~\ref{sec:Conclusion}.
{The Lagrangian of the dark sector is taken to be
\begin{eqnarray}\label{eq:lagrangianHidden}
\mathcal L_h&=&\frac{(\partial S)^2}{2}+\bar\chi\cancel{\partial}\chi\nonumber\\
&-& m_{\chi}\bar\chi \chi-V_S(S)+\bar \chi (\lambda_s+i\gamma_5 \lambda_p)\chi S\,,
\end{eqnarray}}
\noindent
where the dark matter $\chi$ is a Dirac fermion and the light dark component $S$ is assumed to be a real scalar.
{For simplicity, and since it will not affect the rest of our results in what follows, we consider a potential of the form $V_S = m_S^2\frac{S^2}{2}$, which includes only a mass term for $S$. The relevant terms in the Lagrangian for the visible sector can be simply written as
\begin{eqnarray}
\mathcal{L}_{v}&=&\bar N^c \cancel{\partial}N-m_N\bar N^c N+\mathcal L_{\text{SM}} +\hdots \,,
\end{eqnarray}
{which consists of a mass term for the fermion $N$, the Standard Model Lagrangian $\mathcal{L}_{SM}$, and all the interaction terms that enable establishing thermal equilibrium between $N$ and the SM, which are suggested by the ellipses. 
Note that we try to stay as model independent as possible here and do not  specify whether $N$ is part of the SM or if it is included in some BSM sector.} 
As we will see the condition that $N$ is in equilibrium with the SM until the entropy dilution mechanism takes place is the only requirement concerning the visible sector that is necessary for our scenario to be consistent. Further considerations about the visible sector would be model dependent and are out of the scope of our discussion but can be easily {accommodated} to the present discussion.} 

Finally, the Lagrangian encoding the inflaton-portal interaction between the three sectors is given by
\begin{figure}
\begin{center}
\includegraphics[width=0.9\linewidth]{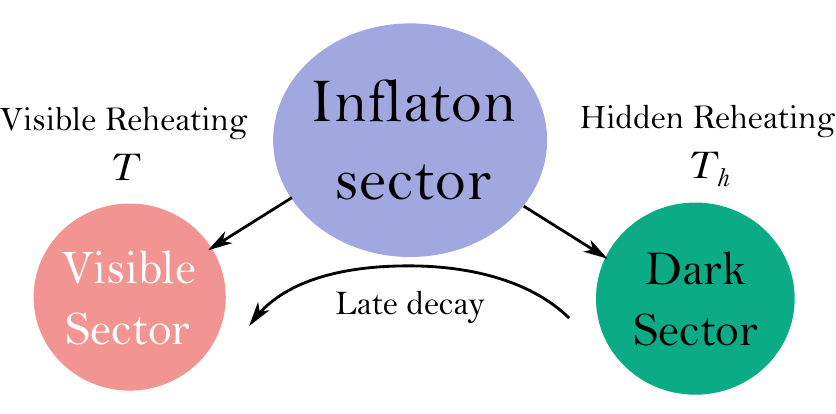}
\caption{\label{fig:scheme}\footnotesize The inflaton portal naturally suppress the decay of the standard model while specifying the reheating processes leading to the hidden and visible sector temperatures.}
\end{center}
\end{figure}
\begin{equation}\label{eq:lagportal}
\mathcal{L}_{portal}=-g_h \phi \bar \chi \chi - g_v \phi N \bar N^c\,.
\end{equation}
{Note that the presence of such coupling between the inflaton, dark matter and the SM may flatten the inflation trajectory significantly at large inflaton field values, and might affect the inflation observables, as we will see in Sec.~\ref{sec:Inflation}.} This Lagrangian will also be responsible for the first stage of reheating of the universe, where the energy of the inflationary sector will be transferred to the visible and hidden sectors, as will be described in Sec.~\ref{sec:Reheating}. 

In order to build a  scenario of highly decoupled sectors, we omitted in the aforementioned Lagrangian several terms which could, in principle, be present in the theory. Let us discuss here the validity or necessity of such assumptions.

Note that the inflaton scalar field, as a singlet under the standard model gauge group, could in principle couple to the higgs boson through operators of the form $ \phi^2 |H|^2$ or $\phi |H|^2$. However, it was shown in \cite{Ema:2017ckf, Ema:2017loe, Enqvist:2016mqj, Gross:2015bea} that such operator could destabilize the Higgs boson during late oscillation of the inflaton and kick the latter away from the electroweak vacuum. Moreover, as we will see in Sec.~\ref{sec:Inflation}, the Yukawa coupling $g_v$ has to be sufficiently small in order not to perturb too much the inflationary trajectory at the loop level. {Therefore, even if such operators would be generated at the loop level  (as it was shown for instance in Ref. \cite{Gross:2015bea} in the case where $N$ would be a right-handed neutrino), the loop corrections to the aforementioned operators would be extremely suppressed for relatively large masses of $N$  (which will be our case).} In this sense, neglecting these couplings as compared to the couplings $g_h$ and $g_v$ that we introduced is technically natural.}

{
As another singlet scalar of the theory, the hidden scalar $S$ could in principle also couple both to the Higgs, the inflaton and to the fermion $N$. For simplicity we assume that there is no mixing term between the inflaton and the hidden scalar. This prevents us from having  to rotate the mass matrix, and work directly in the mass eigenstate basis where the bare quadratic couplings correspond to the physical masses. {In terms of corrections to the lifetime of $S$,} trilinear and quartic couplings of $S$ to the inflaton $\phi$ of the form $S\phi^2$, $S^2\phi^2$ would have no effect beyond introducing loop corrections to the tree-level Lagrangian that would involve more than one inflaton particle. As compared to the tree-level suppression involving only one inflaton propagator, such correction will be irrelevant to what is discussed here. An operator of the form $\phi S^2$ would introduce an additional five dimensional operator of the form $m_{\phi}^{-1}S^2N \overline{N^c}$. Due to the large mass of the inflaton, such an interaction is similar to the interaction through which dark matter annihilates into {a pair of $N$ fermions} through the inflaton portal, and therefore is too feeble to equilibrate the two baths ever. This was already used in \cite{dev, moi} to consider scenarios of dark matter production out of equilibrium. However one should note that the presence of such coupling of the inflaton to the dark scalar might significantly affect the treatment of the reheating since it would increase the amount of energy injected by the inflaton into the dark sector when the inflaton decays. Moreover, non-trivial couplings of the inflaton to the dark scalar might also alter the oscillation of the scalar fields, and the dark scalar might get to oscillate as well after inflation, transferring energy from the inflation sector to the dark sector through classical oscillations. In this paper we assume the inflaton to couple exclusively to fermions in what follows for simplicity and leave the study of multi-scalar interactions for future work.}

{Last but not least, the presence of a significant Yukawa coupling of the scalar $S$ to the {fermions $N$} or any direct coupling between the hidden scalar and the Higgs boson, might lead to undesirable consequences in our scenario. Indeed, such coupling may maintain equilibrium between the hidden and visible baths and have to be forbidden for the whole scenario to be consistent, similarly to Refs. \cite{hooper1, hooper2}. A top-down UV completion of the model would be necessary to motivate such an assumption and generate the appropriate set of couplings considered in this paper. As we will see, assuming the coupling $S N\overline{N^c}$ to be small is technically natural since the loop-induced contributions to such operator are suppressed by the inflaton mass propagator. However such loop induced operator will be the key point of our analysis since the late decay of $S$ will be triggering the entropy-dilution effect in our scenario. Therefore, introducing a bare coupling for such interaction bigger than the loop-induced couplings would require a tuning similar to that which is necessary in \cite{hooper2} in order for the scenario to be successful. We will therefore assume in what follows that no coupling between the visible and the hidden sector is allowed at tree level in our model. We leave for future work the search for an ultra-violet embedding of such theory motivating such strong assumption.}

\section{Hidden and Visible Reheatings}\label{sec:Reheating}

Because the mass of the inflaton ($m_{\phi}\sim 10^{13}$ GeV) is rather large as compared to the masses of the visible- and hidden-sector particles, interactions between the two sectors is extremely suppressed, which prevents the two sectors from equilibrating thermally. Therefore the two sectors will be highly decoupled once they get populated by inflaton decay. Therefore reheating will give rise to two different temperatures $T_h$ and $T$ to the hidden- and visible-sector baths respectively, according to the relative branching ratios of inflaton decay into hidden and visible particles.

The inflaton being the only mediator between the visible sector and the dark sector in our construction, its couplings to the latter are made explicit in the Lagrangian shown in Eq.~\eqref{eq:lagportal} since they will directly play a role in the production of the dark-matter relic abundance.  We would like to stress in this section that the initial ratio of the two associated temperature $T$ and $T_h$ is furthermore fixed by the choice of these couplings, which was not the case in Ref.~\cite{hooper2} where such quantity was arbitrarily chosen. Indeed, demanding that energy be conserved at the reheating time (which is assumed to be instantaneous for simplicity\footnote{Note that treating the reheating as a continuous process can lead to significant corrections of the reheating temperatures \cite{Garcia:2017tuj} but such effect will not affect the initial ratio of temperatures in our case since the interaction of the inflaton to the dark and visible sectors are of similar nature in our toy model. }) yields the relation
\begin{equation}\label{eq:energyCons}
 \rho_h+\rho_v\equiv \mathrm{Br}(\phi\to \chi\chi)\rho_{\phi}+\mathrm{Br}(\phi\to N N)\rho_{\phi}=\rho_{\phi}\,,
\end{equation}
where $\mathrm{Br}(\phi\to \chi\chi)$ and $\mathrm{Br}(\phi\to N N)$ denote the branching ratios of the inflaton decay to the hidden and the visible sectors. These branching fractions can be computed from the Lagrangian in Eq.~\eqref{eq:lagportal}
\begin{equation}
\begin{aligned}\label{eq:branching}
\mathrm{Br}(\phi\to \chi\chi)& =\frac{\Gamma_{\phi\to \chi\chi}}{\Gamma_{\phi}^{tot}}=\frac{g_h^2}{g_h^2+g_v^2}\,,\\
\mathrm{Br}(\phi\to N N)& =\frac{\Gamma_{\phi\to N N}}{\Gamma_{\phi}^{tot}}=\frac{g_v^2}{g_h^2+g_v^2}\,.
\end{aligned}
\end{equation}
At the time of reheating, assuming instantaneous thermalization of the decay products, the hidden sector and visible sector temperatures are given by
\begin{equation} 
T_h^{\text{ inf}} = \left(\frac{30}{\pi^2}\frac{\rho_h}{ g^{\star}_{h\text{, inf}}}\right)^{1/4}\,,\quad
T^{\text{ inf}} = \left(\frac{30}{\pi^2}\frac{\rho_v}{g^{\star}_{\text{inf}}}\right)^{1/4}\,,
\end{equation}
where $g^{\star}_{h,\text{ inf}}$ and $g^{\star}_{\text{ inf}}$ are the number of effective degrees of freedom in the hidden and the visible sector immediately after reheating.  The ratio of the two temperatures at the reheating time is thus given by
\begin{equation}
\xi_{\text{inf}}\equiv \left(\frac{T_h}{T}\right)_{\text{inf}}=\left(\frac{g^{\star}_{\,\text{inf}}}{g^{\star}_{h,\,\text{inf}}}\right)^{1/4}\times\left(\frac{\rho_h}{\rho_v}\right)^{1/4}\,.
\end{equation}
Making use of Eq.~\eqref{eq:energyCons} and \eqref{eq:branching}, one finally obtains
\begin{equation}
\xi_{\text{inf}}=\left(\frac{g^{\star}_{\text{ inf}}}{g^{\star}_{h,\text{ inf}}}\right)^{1/4}\times\left(\frac{g_{h}}{g_{v}}\right)^{1/2}\,.
\end{equation}
One can see at this point why the use of the inflaton portal as a production mechanism for dark matter is of interest in the context of a highly decoupled dark sector: the initial conditions for the hidden- and visible-sector temperatures are entirely specified by the dark-matter production mechanism itself, as opposed to Refs.~\cite{hooper1, hooper2} where the parameter $\xi_\text{inf}$ was a free input of the model.
 
Given entropy conservation in both the hidden and the visible sectors separately, the ratio $\xi(T)$ will evolve with time according to the relation \cite{hooper2}
\begin{eqnarray}
\xi(T)&=&\left(\frac{g_{\star}(T)}{g_{\star\text{ inf}}}\right)^{1/3}\left(\frac{g_{\star\text{ inf}}^h}{g_{\star}^h(T_h)}\right)^{1/3}\xi_{\text{ inf}}\,,\nonumber\\
&=&\left(\frac{g_{\star}(T)}{g_{\star}^h(T_h)}\right)^{1/3}\left(\frac{g^{\star}_{h,\text{ inf}}}{g^{\star}_{\text{ inf}}}\right)^{1/12}\left(\frac{g_{h}}{g_{v}}\right)^{1/2}\,.
\end{eqnarray}
The evolution of $\xi(T)$ is therefore entirely given by $(i)$ the matter content of the model and $(ii)$ the chosen reheating scenario, which, as we will see in the next section, is the key ingredient for producing the right relic abundance in our framework.

{Note that in this section we assumed that the decay of the inflaton is followed by an instantaneous thermalization in both the visible and the hidden sector. Since the couplings we will consider in the dark sector is of order unity, the thermalization of the reheating products in the hidden sector will indeed be fast enough for the instantaneous thermalization assumption to hold. {In the visible sector, as we mentioned earlier, we require that when produced by the decay of the inflaton, the fermions $N$ thermalize instantaneously with the SM. The construction of an explicit scenario for which such assumption would hold is out of the scope of this paper, but can in principle be easily obtained in simple extensions of the SM.}
\section{Dark Matter relic Abundance}\label{sec:RelicDensity}
In the hidden sector, dark matter is maintained in thermal equilibrium with the dark scalar $S$ through the $t$-channel diagram depicted in Fig.~\ref{fig:tchannel}, before the former freezes out thermally.
\begin{figure}
\begin{center}
\includegraphics[width=0.5\linewidth]{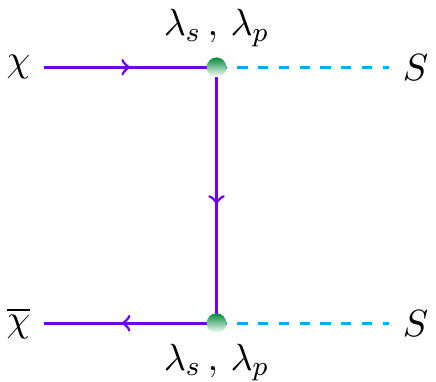}
\caption{\label{fig:tchannel}\footnotesize Annihilation channel of dark matter into dark scalars, which ensures thermal equilibrium in the dark sector before DM freezes out.}
\end{center}
\end{figure}
After dark matter has decoupled,  the remaining population of hidden scalars $S$ may remain in thermal equilibrium before becoming non-relativistic. In the case where equilibrium is maintained for temperatures $T_h\lesssim m_S$, the population may go through a phase of "cannibalism", which may increase the temperature ratio $\xi$ as described in \cite{hooper2}. However it was shown on several examples that the effect of this phase on the  numerical results is as low as $\lesssim$ 5\%. In our case, at tree level, the main annihilation process keeping $S$ in thermal equilibrium is the process $SS\leftrightarrow \chi\chi$ which stops being efficient when dark matter freezes out. We therefore considered for simplicity that the dark scalar also decouples when dark matter freezes out. At the loop level number depleting operators may maintain $S$ in thermal equilibrium for a short time. We checked that the sensitivity of our results to this assumption is however not relevant for the discussion of the dark-matter relic density.

Once the hidden-sector population becomes non-relativistic, it quickly starts dominating the energy density of the universe, provided that it is sufficiently long-lived. The late decay of the hidden scalar will thereafter inject entropy into the visible thermal bath. Let us define $S_i$ to be the entropy of the visible-sector bath at time $t_i$ before any significant fraction of the population of $S$ particles has decayed and $S_f$ to be the entropy of that sector at some time $t_f$ after essentially the entirety of this population has decayed. In the hidden sector, inversely, this decay will have the effect of diluting the dark matter relic density by the ratio $S_i/S_f$ \cite{hooper2}:
\begin{equation}
\left.\Omega_{\chi}h^2\right|_{t_f}=\left.\Omega_{\chi}h^2\right|_{t_i} \times \frac{S_i}{S_f}\,.
\end{equation}
Thus far in the literature, the presence of such a low decay rate for the hidden scalar was introduced by hand with the use of a very small parameter, tuned for the purpose of reproducing the appropriate relic abundance \cite{hooper1,hooper2}. In some of the proposed scenarios, such a tuning was claimed to be technically natural as being protected from large loop corrections \cite{hooper2}. In our case, {the late decay of the hidden scalar $S$, which is assumed to have no tree-level contact interactions with the SM, is generated at the loop level and suppressed by the inflaton propagator $\Delta_{\phi}\sim m_{\phi}^{-2}$}, as can be seen in Fig.~\ref{fig:decay} where a loop of dark-matter particles exchange an inflaton with a pair of {fermions $N$} which are assumed to be in equilibrium with or to decay rapidly to SM particles thereafter.
\begin{figure}
\begin{center}
\includegraphics[width=0.9\linewidth]{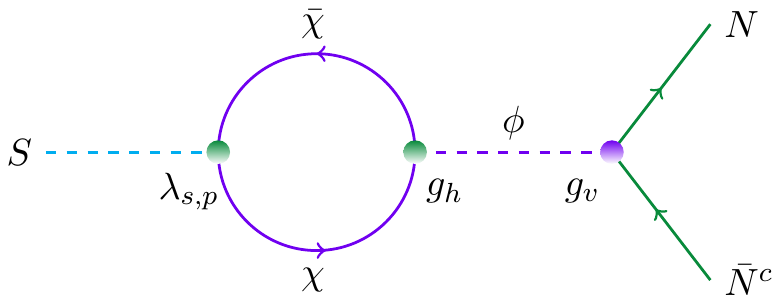}
\caption{\label{fig:decay}\footnotesize Decay channel of the hidden scalar $S$  --- through a loop of dark matter particles and the exchange of an inflaton $\phi$ --- into $N$ fermions.}
\end{center}
\end{figure}
 The late decays of hidden scalars into SM particles could however destroy Big Bang Nucleosynthesis (BBN) predictions if these decays reheat the universe at a temperature as low as $\lesssim 10 \mathrm{MeV}$. Therefore we require that the hidden scalar lifetime $\tau_S$ satisfies the BBN bound
\begin{equation}\label{eq:BBNbound}
T_\text{RH}\equiv\left(\frac{90}{8\pi^3g_{\star}}\right)^{1/4}\sqrt{\frac{M_p}{\tau_s}}>10~\mathrm{MeV}\,.
\end{equation}
In our specific model the decay rate is given by
\begin{equation}
\Gamma_{S}=\frac{\lambda_s^2+\lambda_p^2}{8\pi}\left(\frac{g_h g_v  }{8\pi^2}\frac{m_{\chi}^2}{m_{\phi}^2}\right)^2 m_{S}\left(1-4\frac{m_{N}^2}{m_{S}^2}\right)^{3/2}\,, 
\end{equation}
and the reheating temperature can be estimated as
\begin{eqnarray}\label{eq:decay}
T_\text{RH}&\approx& 119~\mathrm{MeV}\left(\frac{10^{13}\mathrm{GeV}}{m_{\phi}}\right)^2\left(\frac{m_{\chi}}{50\mathrm{PeV}}\right)^{5/2}\nonumber\\
&\times& \left(\frac{m_{\chi}/m_S}{20}\right)^{-1/2}g_h g_v\sqrt{\frac{\lambda_s^2+\lambda_p^2}{2}}\,.
\end{eqnarray}
For an inflaton mass of $10^{13}~\mathrm{GeV}$, {and masses in the dark sector of order $\mathcal{O}(10-100)$ PeV as obtained in \cite{hooper2},} the universe is reheated to a low temperature close to the BBN bound \eqref{eq:BBNbound}, which is necessary for the entropy-dilution mechanism to work efficiently.

{An analytic calculation of the relic abundance is given in Appendix \ref{sec:RelicAbundance} in the case in which the annihilation cross section for $\chi$ is predominately $s$-wave. In the situation where our couplings $\lambda_s$ and $\lambda_p$ both take non-vanishing values this will indeed be the case, as discussed in Appendix \ref{AppendixA}.}
\begin{figure*}[t!]
\begin{subfigure}[b]{0.48\textwidth}
        \centering
        \includegraphics[width=\linewidth]{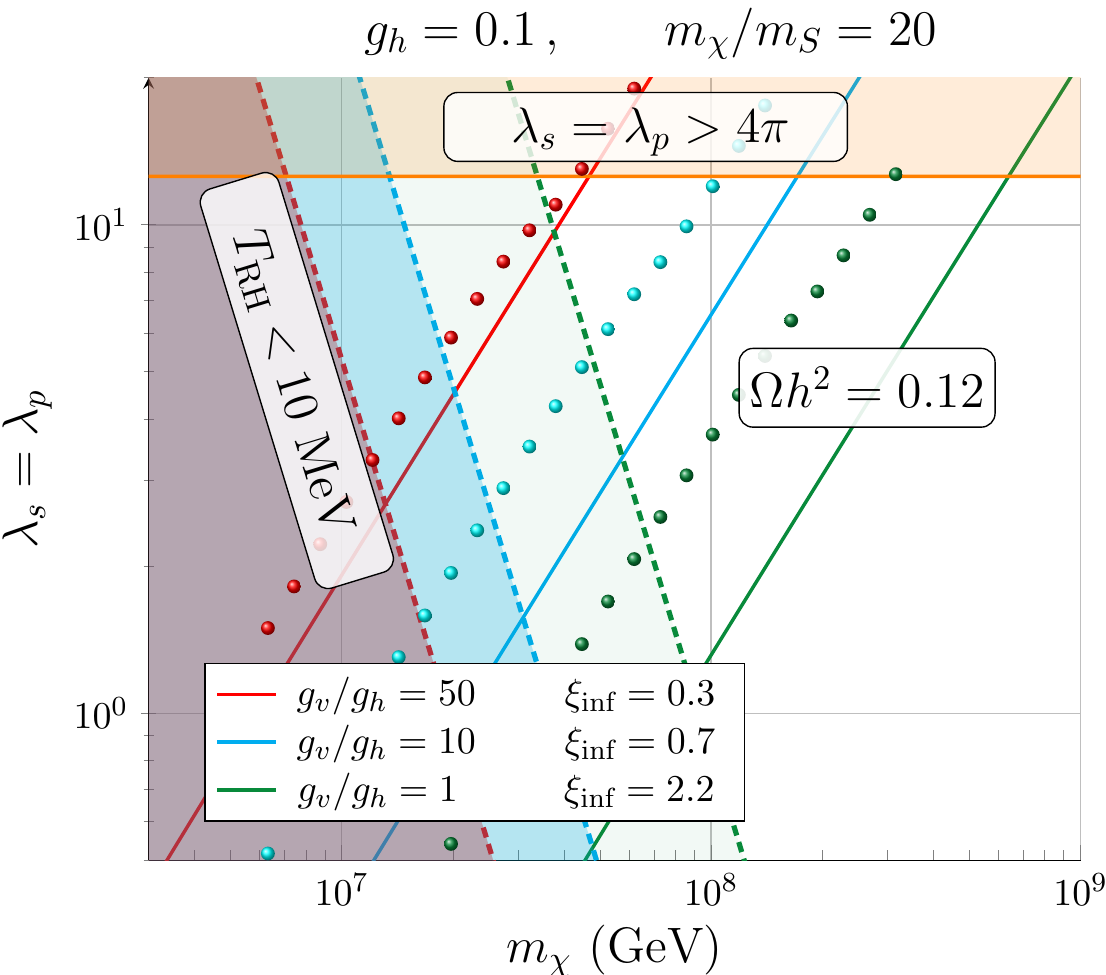}
\end{subfigure}\hspace{0.02\linewidth}
\begin{subfigure}[b]{0.48\textwidth}
        \centering
        \includegraphics[width=\linewidth]{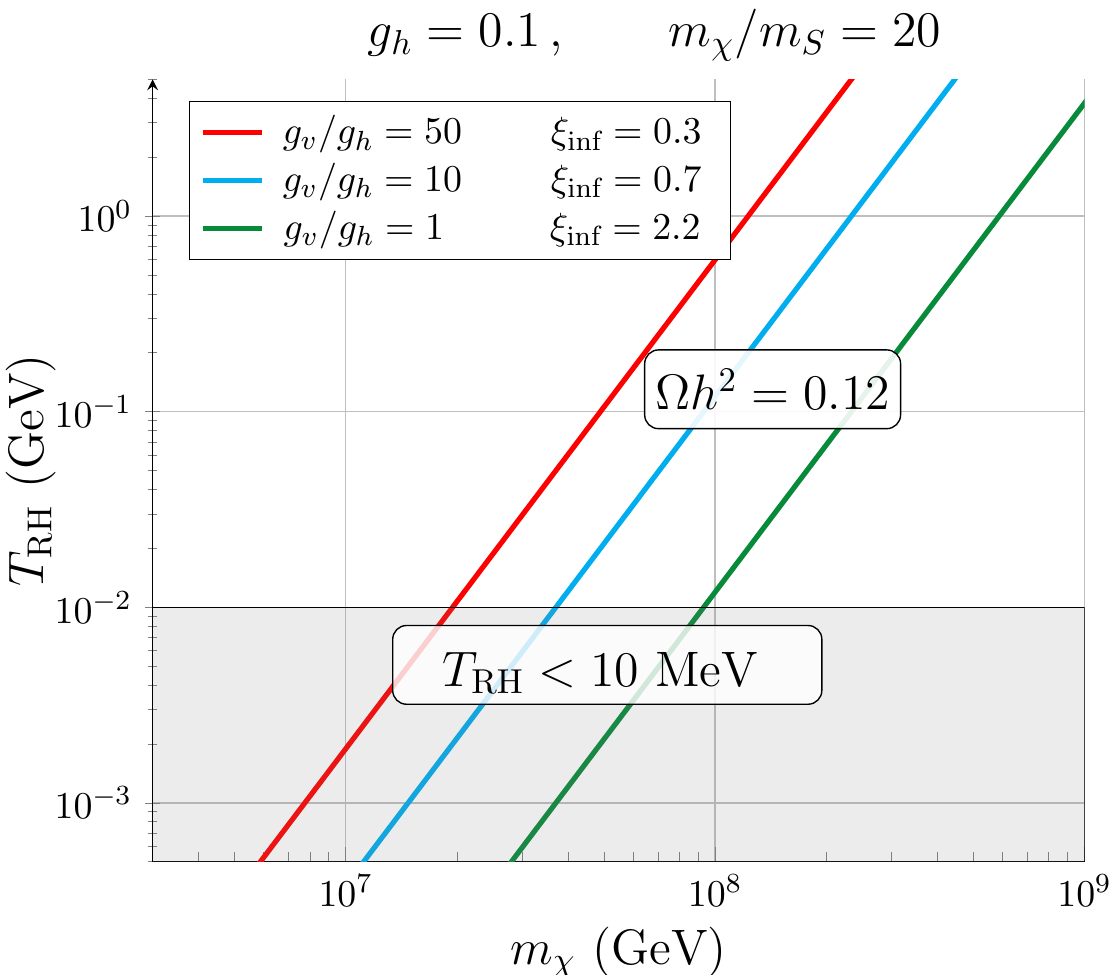}
\end{subfigure}
\caption{\label{fig:Relic}\footnotesize $(Left)$ Parameters leading to a relic density of dark matter $\Omega_{ \chi}h^2=0.12$ fixing $g_h=0.1$ and varying $g_v$ and for a mass ratio $m_{\chi}/m_S=20$. $(Right)$ Reheating temperature as a function of the dark matter mass for similar sets of parameters.}
\end{figure*}
{Defining an effective coupling constant $\alpha_{\chi}$ by
\begin{equation}
\langle\sigma v\rangle_{s-wave}\equiv 2\pi\frac{\alpha_{\chi}^2}{m_{\chi}^2}\,,
\end{equation}
and distinguishing the two limit cases where the effective number of relativistic degrees of freedom $g_{\star}^\text{eff}=g_{\star}+g_{\star}^h\xi_\text{FO}^4$ is overwhelmingly dominated at freeze-out by the visible sector contribution $g_{\star}$ or by the hidden sector one $g_{\star}^h\xi_\text{FO}^4$, one obtains the analytic estimation for the relic density, 
\begin{eqnarray}
&&\bullet \quad g_{\star}^h\xi_\text{FO}^4\lesssim g_{\star} :\nonumber\\
&&\frac{\Omega_{\chi}h^2}{S_f/S_i}\approx 0.16\left[1+0.04\ln\left(\frac{\xi_\text{FO}^2}{\sqrt{g_{\star}+\xi_\text{FO}^4 g_{\star}^h}}\frac{\alpha_{\chi}^2}{0.8^2}\frac{50\mathrm{PeV}}{m_{\chi}}\right)\right]\nonumber \\
&&\times\xi_\text{FO}^{-2} \left(\frac{T_\text{RH}}{10\mathrm{MeV}}\right)\left(\frac{m_{\chi}}{50\mathrm{PeV}}\right)\left(\frac{m_{\chi}/m_S}{20}\right)\left(\frac{0.8}{\alpha_{\chi}}\right)^2\,.\label{eq:Relic1}\\
&&\bullet \quad g_{\star}^h\xi_\text{FO}^4\gtrsim g_{\star} :\nonumber\\
&&\frac{\Omega_{\chi}h^2}{S_f/S_i}\approx \left.\left(\frac{\Omega_{\chi}h^2}{S_f/S_i}\right)\right|_{g_{\star}^h\xi_\text{FO}^4\lesssim g_{\star}}\times \sqrt{\frac{g_v}{g_h}}\left(2-\sqrt{\frac{g_v}{g_h}}\right)\,,\label{eq:Relic2}
\end{eqnarray}
where, at freeze-out, the value of $\xi$ can be estimated to be (see Appendix \ref{sec:RelicAbundance})
\begin{equation}
\xi_\text{FO}=\left(1+\frac{c_{\chi}g_{\chi}}{c_S g_S}\right)^{1/3}\left(\frac{g^{\star}_{\text{ inf}}}{g^{\star}_{h,\text{ inf}}}\right)^{1/4}\times\left(\frac{g_{h}}{g_{v}}\right)^{1/2}\,,
\end{equation}
with $c_{S,\chi}=1~(7/8)$ for bosonic (fermionic) $S$ and $\chi$ and $g_{S}=1$, $g_{\chi}=4$ being the numbers of internal degrees of freedom.
}
Note that these relations are independent of the specific formulas for the decay rate and the annihilation cross section. In the particular model we consider here, the decay rate is given by Eq.~\eqref{eq:decay} and the $s$-wave part of the annihilation cross section is given in Appendix \ref{AppendixA}, with $\alpha_\chi$ given by
\begin{equation}\label{eq:alpha}
\alpha_{\chi}\equiv\frac{\lambda_s\lambda_p}{4\pi}\,.
\end{equation}
{In what follows we will for simplicity consider the case $\lambda_s=\lambda_p\equiv\lambda_\chi$\footnote{Suppressing one or the other coupling as compared to its counterpart would have the effect of rendering the annihilation process $p$-wave suppressed, therefore increasing the dark-matter relic density and rendering the overclosure of the universe more difficult to avoid without destroying the BBN predictions.}.}

The Eqs.~\eqref{eq:Relic1}-\eqref{eq:Relic2} constitute one of the main results of this paper. Indeed one can clearly see that if one requires
{
\begin{itemize}
\item[{\em (i)}] the couplings $\lambda_\chi$, $g_v$ and $g_h$ to be  $\lesssim 1$,
\item[{\em (ii)}] the inflaton mass to be of order $m_{\phi}\sim 10^{13}~\mathrm{GeV}$, as required in the case of large-field inflation scenarios, 
\item[{\em (iii)}] the dark-matter mass to be heavier than $\gtrsim$ 10 PeV --- which is then necessary for satisfying the BBN bound in Eq.~\eqref{eq:BBNbound},
\item[{\em (iv)}] the mass hierarchy in the dark sector to be reasonably small $m_{\chi}/m_S \gtrsim$ $\mathcal O(1-10)$ --- otherwise the perturbative regime is as well excluded by the BBN bound,
\end{itemize}
then the dark-matter relic abundance accords with its measured value. The inflaton mass plays a key role in our analysis and requiring that $m_\phi$ falls within the range of values favoured by astrophysical observations in the context of large-field inflation models (see e.g. \cite{Linde:2014nna}) severely constrains the allowed range for the dark-matter mass. In particular we will see that it imposes a lower bound of roughly $m_{\chi}\gtrsim \mathcal O(10)~\mathrm{EeV}$. In Refs. \cite{hooper1,hooper2}, the inflaton is implicitly present but only plays the role of producing the visible and dark baths. However, in our paper, the inflaton is also a very convenient mediator between the dark sector and the visible sector. Its presence highly suppresses the decay of the dark scalar, which allows the entropy-dilution mechanism to work successfully without the need of any new physics apart from the inflationary sector to relate the visible and invisible baths. Interestingly, the temperature ratio $\xi_\text{inf}$ after reheating and the decay rate of $S$, which were two separate ingredients in Refs.~\cite{hooper1,hooper2} are now intimately related.} Indeed, both quantities depend on the couplings of the inflaton to the visible and dark sectors. We will see in what follows that this relation constrains considerably the parameter space.

We numerically solved the Boltzmann equations from the reheating time (assumed to take place at energies higher than the dark-matter mass $T_\text{RH}\gtrsim m_{\chi}$) to present time, tracking the relic abundance and its dilution due to the transfer of entropy from the hidden bath to the visible thermal bath. For simplicity we assume the pseudoscalar and scalar couplings $\lambda_p$ and $\lambda_s$ to be equal and explore different hierarchies between the couplings $g_h$ and $g_v$. Different choices of the ratio $g_h/g_v$ will lead to different initial conditions for the temperature ratio $\xi_\text{inf}$.

The results are plotted in the left panel of Fig.~\ref{fig:Relic} in the plane $(m_{\chi},\lambda_s=\lambda_v)$ where we fix the coupling $g_h=0.1$, the mass ratio $m_{\chi}/m_S=20$ and allow the coupling of the inflaton to the visible sector $g_v$ to vary. Dashed lines represent the BBN bound and the orange region is the non-perturbative limit. The solid lines stand for the parameter satisfying $\Omega h^2=0.12$ according to Eq.~\eqref{eq:Relic1} whereas the dots represent the results obtained by numerically solving the Boltzmann equations. As expected from the BBN bound in Eq.~\eqref{eq:BBNbound} the dark-matter mass is generically constrained to be  larger than $\mathcal O(10)$ PeV in order for the relic density to agree with observation while restricting all couplings to the perturbative range $g_h,g_v,\lambda_s,\lambda_p\ll 4 \pi$. {In the right panel of Fig.~\ref{fig:Relic} we indicate the temperature at which the decay of $S$ reheats the visible thermal bath by copiously { producing pairs of $N$ fermions}. One can observe that the reheating temperature is bounded from above, due to the perturbativity bound and the BBN bound indicated in the left panel, and lies in the range $10\mbox{~MeV}\lesssim T_\text{RH}\lesssim \mathrm{GeV}$ for a dark-matter candidate with a mass of order \mbox{10 PeV - EeV}.}

{In order for $S$ to decay directly into a pair of fermions $N$, the latter have to be light enough $m_N<m_S/2$ for this decay channel to be kinematically open. {In this analysis we have taken the mass of the fermion $N$ to be negligible compared to the {mass of $S$ ($m_N~\ll~m_S$) such that it can be considered massless in the analysis.} One could wonder whether the production of these fermions would immediately be followed by their decay into visible particles. A later decay of $N$ could indeed violate the BBN bound that we have imposed. Again, we will assume that either $N$ instantaneously equilibrate with the SM at the time of entropy dilution {or that it decays} instantaneously into SM states.}

{Before we continue, one should note that introducing couplings of $\mathcal O(1)$ of the inflaton to fermions could alter the present discussion in two ways: On the one hand, fermionic couplings of the inflaton to the matter sector allow the scalar mass to run with the energy scale, altering the shape of the inflationary potential through loop corrections. Such corrections will be introduced and studied in details in the next section. On the other hand, whereas parametric resonances of the particle production rate at the end of inflation are expected to take place whenever the inflaton would couple to scalar particles, it was shown in Refs.~\cite{Greene:2000ew,Greene:1998nh} that a parametrically resonant production of fermions could happen as well whenever the Yukawa couplings are too large. As we will see in what follows, the presence of radiative corrections to the inflation trajectory will actually require the Yukawa couplings of our model to be smaller than $10^{-6}-10^{-4}$ in order for the scenario to be experimentally viable. In Refs.~\cite{Greene:2000ew,Greene:1998nh} a sufficient condition is that the Yukawas don't exceed $10^{-6}$ which could be in tension with our perturbative treatement of the reheating in our model. However such constraint strongly depends on the inflation potential chosen, and in particular on the inflaton mass and field value at the end of inflation. In our scenario the inflaton field value at the end of inflation is typically sub-Planckian, which differs from the quadratic and quartic cases studied in Refs.~\cite{Greene:2000ew,Greene:1998nh}, therefore possibly altering such constraint. A dedicated study of the fermionic reheating in the context of alpha-attractor models would be required to properly address this question, and we shall leave this for future investigations.}

\section{Inflationary Trajectory, Backreaction}
\label{sec:Inflation}
Until now we have {not specified the explicit form of the inflation potential $V_\text{inf}(\phi)$. We will now introduce explicit inflation scenarios and explore how the inflationary {sector actually} turns out to be the main source of {observational} constraints in the following sections.}

{In order to study the inflationary dynamics in various cases we will in what follows focus on a class of inflationary models called $\alpha$-attractors (see e.g. \cite{Kallosh:2015lwa}) in its E-model version. Such scenario corresponds to a potential of the form
\begin{equation}
V(\phi)=V_0\left( 1- e^{-\sqrt{\frac{2}{3\alpha}}\frac{\phi}{M_{p}}}\right)^2\,.
\end{equation}
In the vacuum, the mass of the inflaton $m_{\phi}$ can be expressed as
\begin{equation}
m_{\phi}^2=\frac{4 V_0}{3 \alpha M_{p}^2}\,.
\end{equation}
In the limit of large $\alpha$ the potential is nearly quadratic and one recovers the standard chaotic inflation
\begin{equation}
V_\text{inf}^{\alpha\gg 1}(\phi)\approx\frac{m_{\phi}^2}{2}\phi^2\,.\quad \text{(Chaotic)}
\end{equation}
In the specific case of $\alpha=1$ one obtains the Starobinsky inflation potential \cite{Starobinsky:1983zz}
\begin{equation}
V_\text{inf}^{\alpha=1}(\phi)\approx V_0\left( 1- e^{-\sqrt{\frac{2}{3}}\frac{\phi}{M_{p}}}\right)^2\,.\quad \text{(Starobinsky)}
\end{equation}
Lower values of $\alpha$ are usually inspired by supergravity models. $\alpha=1/3$ would naturally be expected from {\mbox{$\mathcal N=4$}} SUGRA \cite{Cremmer:1977tt,Bergshoeff:1980is,deRoo:1984zyh,Ferrara:2012ui} while $\alpha=1/9$ are typically obtained in so-called GL models \cite{Goncharov:1985yu,Goncharov:1983mw,Linde:2014hfa}. 
Other inflation scenarios could of course be considered but we will in this work focus on the two extreme cases {--- chaotic inflation and Starobinsky inflation}, using {$\alpha=1$ and the limit $\alpha\gg 1$.}
We checked that the choice  $\alpha=1$ rather than $\alpha=1/3$ or $\alpha=1/9$ has no significant effect on the conclusions of our work since models with small values of $\alpha$ give very similar predictions in terms of the tensor-to-scalar ratio and spectral index measurement.
}

{Generically, for such potentials, the normalization of the power spectrum of scalar perturbations fixes \mbox{$m_{\phi}\sim 10^{13}$ GeV} in all the different cases.} It is well known that such a scalar potentials will receive radiative corrections as soon as the inflationary sector is coupled to any matter sector\footnote{Note that in supersymmetric theories, such corrections do not arise.}\cite{NeferSenoguz:2008nn}. In our scenario, the coupling of the inflaton both to dark matter and to the SM fermion $N$ will modify the potential at high energy:
\begin{equation}\label{eq:radiative}
V(\phi)=V_\text{inf}(\phi)+V_\text{1-loop}\,,
\end{equation}
where
\begin{eqnarray}\label{eq:PotLoop}
V_\text{1-loop}&=&-\frac{1}{32\pi^2}\left\{\left(g_v \phi + m_N\right)^4\ln\left[\frac{(g_v \phi+m_N)^2}{\mu^2}\right]\right.\nonumber\\
&+&\left.\left(g_h \phi + m_{\chi}\right)^4\ln\left[\frac{(g_h \phi+m_{\chi})^2}{\mu^2}\right]\right\}\nonumber\\
&\approx &-\frac{1}{32\pi^2}\left[g_v^4 \phi^4 \ln\left(\frac{g_v^2 \phi^2}{\mu^2}\right)+g_h^4 \phi^4 \ln\left(\frac{g_h^2 \phi^2}{\mu^2}\right)\right]\,,\nonumber\\
\end{eqnarray}
where $\mu$ is the renormalization scale. In going from the first to the second equality in Eq.~\eqref{eq:PotLoop}, we have assumed that $m_N,m_{\chi}\ll g_h \phi, g_v\phi$. This follows from the fact that on the one hand $\phi\gtrsim M_p$ during inflation, while on the other hand the couplings $g_{v,h}$ are required to be quite large in order to be consistent with experimental bounds, as we discuss in more detail below. It is therefore clear that the radiative corrections arising from fermion loops will flatten the potential for large field values of the inflaton, as long as the couplings are sufficiently large. As already suggested in Refs.~\cite{NeferSenoguz:2008nn, Buchmuller:2015oma}, such a flattening can be dramatic. It may destabilize the potential around the Planck scale and therefore destroy the slow roll inflation scenario. However, it may provide a method for easing tensions between chaotic inflation and the constraints established by recent measurements of the CMB \cite{Ade:2015xua}. In what follows we will fix the renormalization scale to be $\mu=M_p$.

Typically, large values of the couplings $g_h$ and $g_v$ will have the tendency of adding large quartic corrections to the inflaton potential. Such corrections can either destabilize the inflationary trajectory or simply destroy the slow-roll regime. In the case of chaotic inflation, such corrections have the effect of steepening the inflaton potential at large field values. As far as observational constraints are concerned, this has a dramatic consequence on the tensor-to-scalar ratio and the spectral index predicted by the theory. Therefore the case of chaotic inflation cannot be accommodated in our framework. Note that adding a negative quartic  counter term to the potential could turn such steepening into a flattening of the trajectory, which could help  alleviate the discrepancy between the prediction for the observables and the Planck constraints. However, the adjustment required to do so would render the model extremely fine-tuned. We will therefore ignore such possibility in what follows and focus on the case $\alpha=1$, which corresponds to Starobinsky inflation. Note that a different choice of $\alpha\sim \mathcal O(1)$ would not alter the main conclusions of our work.


As we will see in the next sections, the values of the couplings $g_v$ and $g_h$ which are acceptable such that the inflation trajectory is not destabilized are typically smaller than $10^{-3}$. The relic density and BBN constraints for such low values of the couplings are indicated in Fig.~\ref{fig:observables} for a mass ratio $m_{\chi}/m_S=2$ an $g_{h},g_v\sim 10^{-3}$ one can see that the preferred region for the dark matter mass is in the range $m_{\chi}\sim\mathcal{O}(\mathrm{2-20})\mathrm{EeV}$. An interesting feature of such result is that the choice of the mass ratio $m_{\chi}/m_S$ does not influence this lower bound of 2 EeV for the dark-matter mass while it does impact the upper bound. To understand this, we refer the reader to the BBN bound in Eq.~\eqref{eq:BBNbound}. The lower bound for the dark-matter mass corresponds to the point where $T_\text{RH}\sim 10~\mathrm{MeV}$. For a fixed reheating temperature, one can invert Eq.\eqref{eq:BBNbound} for the the coupling $\lambda_s=\lambda_p$ to find that it scales like $\lambda_s\propto \sqrt{(m_{\chi}/m_S)/m_{\chi}^5}$. It is then easy to see from Eq.~\eqref{eq:Relic1} that the relic density does not depend on the ratio $m_{\chi}/m_S$, but only on the dark-matter mass.

\begin{figure}
\begin{center}
\includegraphics[width=\linewidth]{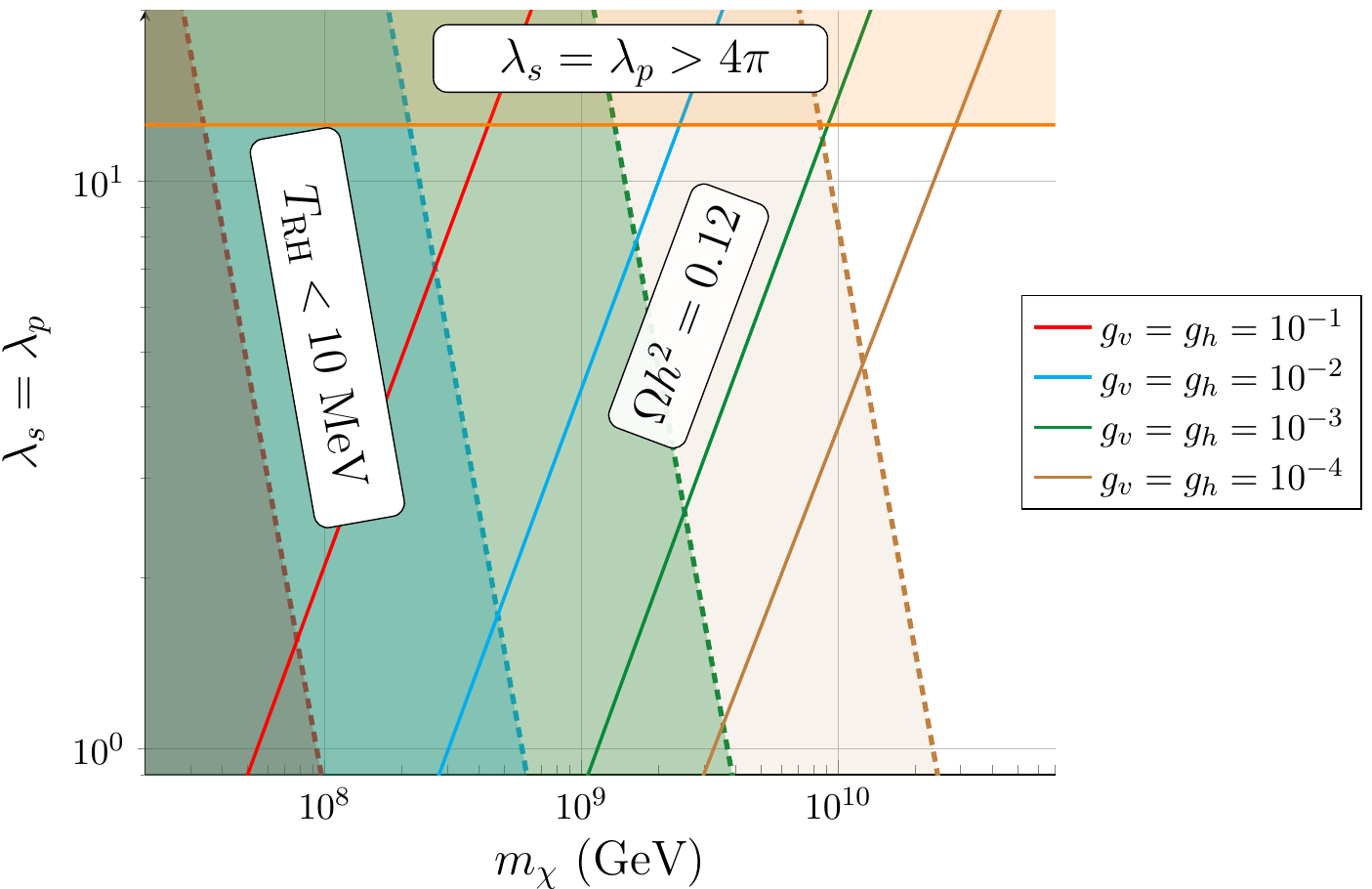}
\caption{\label{fig:observables} Relic density constraints (plain lines), BBN constraints (dashed lines) and perturbativity limit (orange shaded region) for equally small values of the couplings $g_h=g_v$ and for a mass ratio $m_{\chi}/m_S=20$.}
\end{center}
\end{figure}
{
\section{Number of E-folds and Early Matter Domination Era}
\label{sec:EMD}
As we have seen in the previous section, the presence of couplings between the inflaton and the matter sector unavoidably leads to a backreaction on the inflation trajectory through loop corrections, and can potentially be dangerous, because it destabilizes the inflationary trajectory. Such destabilization typically arises at large values of the field at which the inflation observables are computed, which corresponds to horizon crossing of the pivot mode $k_\star=0.05~\mathrm{Mpc}^{-1}$ used by the Planck collaboration~\cite{Akrami:2018odb}.

Depending on the value of the number of e-folds $N_\star$ --- required between the end of inflation and the moment where the pivot mode studied by Planck crosses the horizon --- {the value of the observables} can be drastically different and a given model of inflation can be totally ruled out by present measurements in the case where $N_\star$ is too small.}

{In practice, the number of e-folds of expansion which were present between horizon crossing of the pivot mode $k_\star$ and present times depends on the {cosmological history} and is constrained {by observations}. In the standard thermal scenario, 
{inflation is followed subsequently by a short effective period of matter domination (which corresponds to the oscillation of the inflaton about its minimum before it decays), a phase of reheating (which corresponds to the inflaton decay and thermalization of its decay products) and the usual radiation-dominated, matter-dominated era, and finally dark-energy dominated era.}
Depending on how late the reheating happens, some e-folds of expansion can take place between the end of inflation and 
 the reheating phase~\cite{Ueno:2016dim,Ellis:2015pla}. Depending on the reheating temperature, the number of e-folds might therefore vary between $N_\star~=~50$ and $N_\star~=~60$ in standard cosmological scenarios.
In the literature, these are the usual benchmark points studied in order to derive constraints on the inflationary sector. 
In our case, the {expansion history of the universe} is non-standard and in particular involves an early period of matter domination (EMD) {which needs to be traced in our numerical simulations since it affects the precise number of e-folds of inflation.}}

{Given the pivot mode $k_\star$, one can compute the number of inflationary e-folds $N_\star$ which take place between horizon crossing ($k_\star=a_\star H_\star$) and the end of inflation using the relation~\cite{Liddle:2003as,Allahverdi:2018iod,Easther:2013nga,Das:2015uwa,Kofman:1994rk,DiMarco:2018bnw}
\begin{equation}
\frac{k_{\star}}{a_0 H_0}=e^{-N_{\star}}\frac{a_\text{end}}{a_\text{reh}}\frac{a_\text{reh}}{a_\text{EMD}}\frac{a_\text{EMD}}{a_\text{dec}}\frac{a_\text{dec}}{a_\text{eq}}\frac{H_{\star}}{H_\text{eq}}\frac{a_\text{eq}H_\text{eq}}{a_0 H_0}\,.
\end{equation} 
In this equation, quantities with the subscript {``0''} correspond to their value at the present day, whereas {``end''} stands for the end of inflation, {``reh''} for the reheating time corresponding to the inflaton decay, {``EMD''} refers to the beginning of the {early matter-dominated} period, {``dec''} corresponds to the moment {when} the dark scalar decays {and} the dark-matter relic density gets diluted and {``eq''} refers to the time of the matter-radiation equality. Taking the logarithm of this expression, and using the fact that radiation and matter particle baths evolve as $\propto T^4$ and $T^3$ respectively, one obtains
\begin{eqnarray}\label{eq:efolds}
N_\star &=&-\ln \frac{k_\star}{a_0 H_0}+\frac{1}{3}\ln\frac{\rho_\text{reh}}{\rho_\text{end}}+\frac{1}{4}\ln\frac{\rho_\text{EMD}}{\rho_\text{reh}}\nonumber\\&+&\frac{1}{3}\ln\frac{\rho_\text{dec}}{\rho_\text{EMD}}+\frac{1}{4}\ln\frac{\rho_\text{eq}}{\rho_\text{dec}}+\ln \frac{H_\star}{H_\text{eq}}+\ln\frac{a_\text{eq}H_\text{eq}}{a_0 H_0}\,.\nonumber\\
\end{eqnarray}
In this equation, $H_\star$ can be related to $\phi_\star$ using the slow roll approximation
\begin{equation}
H_\star^2\approx \frac{V_\text{inf}(\phi_\star)}{3 M_p^2}\,,
\end{equation}
where $\phi_\star$ is the value of the field, $N_\star$ e-folds before the end of inflation. Therefore Eq.~\eqref{eq:efolds} is an implicit equation in $N_\star$, which can be solved numerically. The values of the different energy densities exhibited in Eq.~\eqref{eq:efolds} are detailed in Appendix~\ref{sec:efolds}. }

{In order to probe the parameter space, we consider benchmark values for the ratio $m_{\chi}/m_S=5,20$ and $g_v/g_h=1,5$ and scan over the value of the dark-matter mass $m_{\chi}$ and the parameter $g_h$. For every point of the scan, we compute the mass of the inflaton necessary to satisfy {the normalization constraint $m_{\phi}~\approx~10^{13} \rm GeV$}, and then adjust the value of the dark coupling $\lambda_{s}=\lambda_p$ in order for the dark-matter relic density to {match observations}. Given these fixed parameters, we thus solve {numerically} Eq.~\eqref{eq:efolds} in order to find the number of e-folds $N_\star$. We additionally compute the tensor-to-scalar ratio and spectral index predicted by the theory and confront it to the most recent Planck results.}

{As a result,  the number of e-folds predicted for any single point of the parameters space which are not excluded by the BBN bound is constrained to be roughly $N_\star\lesssim 50$. Our results are presented in Fig.~\ref{fig:money} in which all the different constraints we {have explored} are taken into account. The violet region excludes the non-perturbative region of the parameter space $\lambda_\chi\leqslant 4\pi$, while the gray-shaded area excludes the parameters for which $T_\text{RH}<10~\mathrm{MeV}$. The ``No Inflation" region corresponds to situations for which the radiative corrections to the inflaton potential start dominating over the tree-level inflationary trajectory. This corresponds to the case where $g_h, g_v \gtrsim 5\times 10^{-4}$. Therefore quartic contribution destroy the slow-roll regime for large inflation field values. Finally, the constraints on the tensor-to-scalar ratio and spectral index are indicated by the brown-shaded regions, which represent the $1\sigma$ exclusion limit published by the Planck collaboration~\cite{Akrami:2018odb}.}

{One can clearly see that the parameter space shrinks for large values of the ratios $m_\chi/m_S,\, g_v/g_h\gg 1$. 
In the case where $g_v/g_h\ll 1$, the relic density formula has to be taken from Eq.~\eqref{eq:Relic2} in which the factor $\sqrt{g_v/g_h}$ naturally reduces the relic density value. It is therefore easier to obtain the correct value of the dark-matter relic abundance for reasonable values of $\lambda_\chi$. However, as can be seen from the top panels of Fig.~\ref{fig:money}, the constraints provided by the Planck collaboration for inflation observables turn out to be more stringent than in the other cases. This is due to a non-trivial tension between the loop correction to the inflation potential which tend to destroy the slow-roll regime for too large values of $g_h$ and $g_v$ and the change of the number of e-folds imposed by Eq.~\eqref{eq:efolds}, which eliminates solutions where these couplings are too large. Therefore the favoured parameter space is clearly}
\begin{equation}
m_\chi\gtrsim 10~\mathrm{EeV}\,,\quad m_S\lesssim m_\chi\,,\quad g_h\gtrsim g_v\,.
\end{equation}
{Better constraints on inflation, and especially on the spectral index (since it is the major source of constraint in the case of plateau-inflation scenarios) {will allow us to put more stringent constraints or even rule out completely the parameter space} in future years.}

\begin{figure*}
\vspace{-0.7cm}
\begin{subfigure}[b]{0.475\textwidth}
        \centering
        \includegraphics[width=\linewidth]{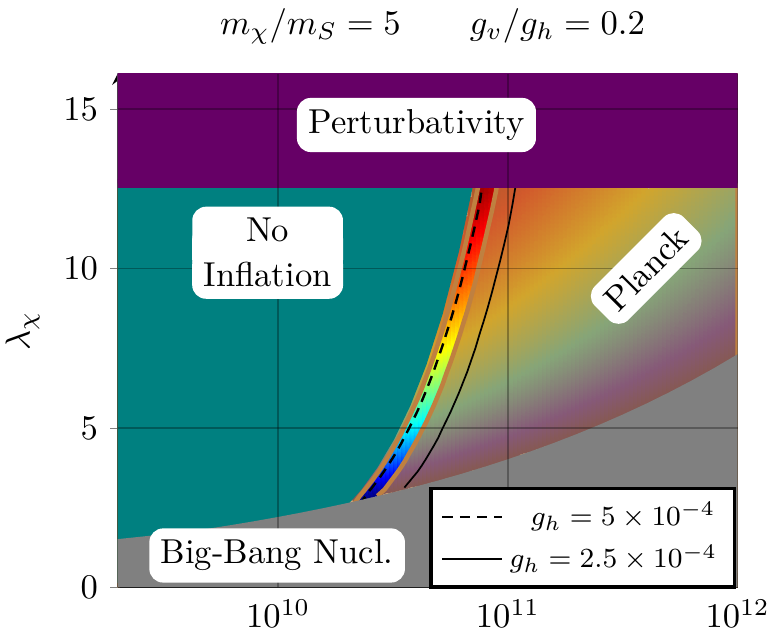}
\end{subfigure}
\begin{subfigure}[b]{0.437\textwidth}
        \centering
        \includegraphics[width=\linewidth]{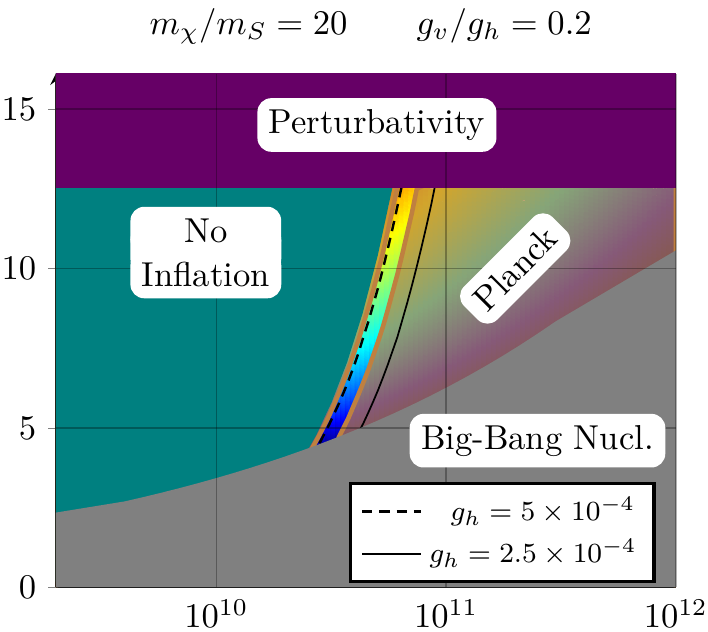}
\end{subfigure}\\
\vspace{0.3cm}
\begin{subfigure}[b]{0.475\textwidth}
        \centering
        \includegraphics[width=\linewidth]{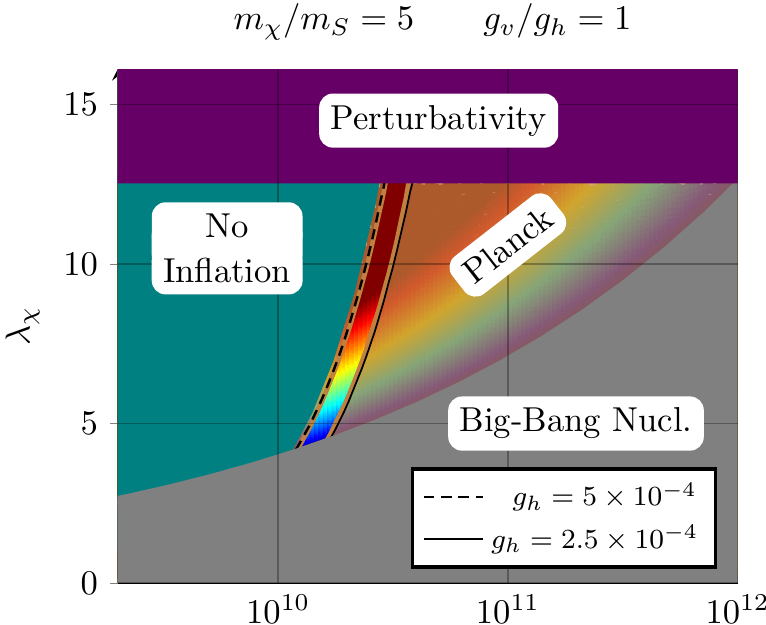}
\end{subfigure}
\begin{subfigure}[b]{0.437\textwidth}
        \centering
        \includegraphics[width=\linewidth]{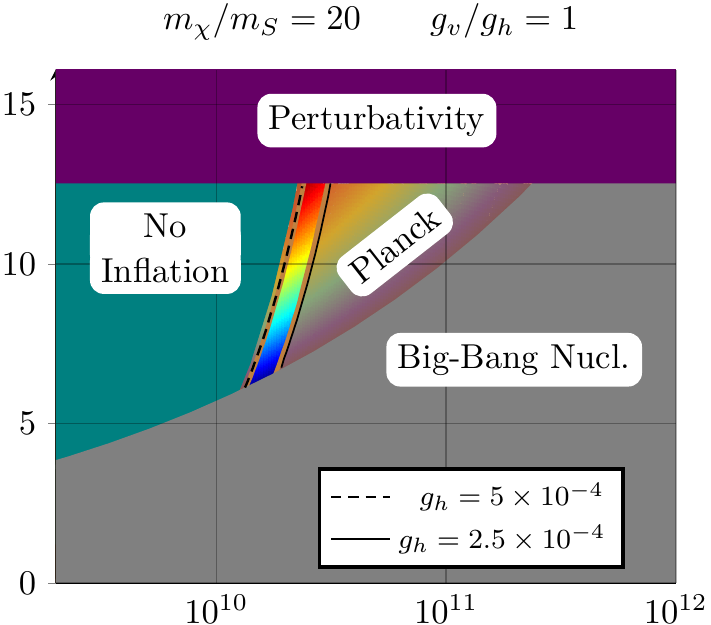}
\end{subfigure}\\
\vspace{0.3cm}
\begin{subfigure}[b]{0.475\textwidth}
        \centering
        \includegraphics[width=\linewidth]{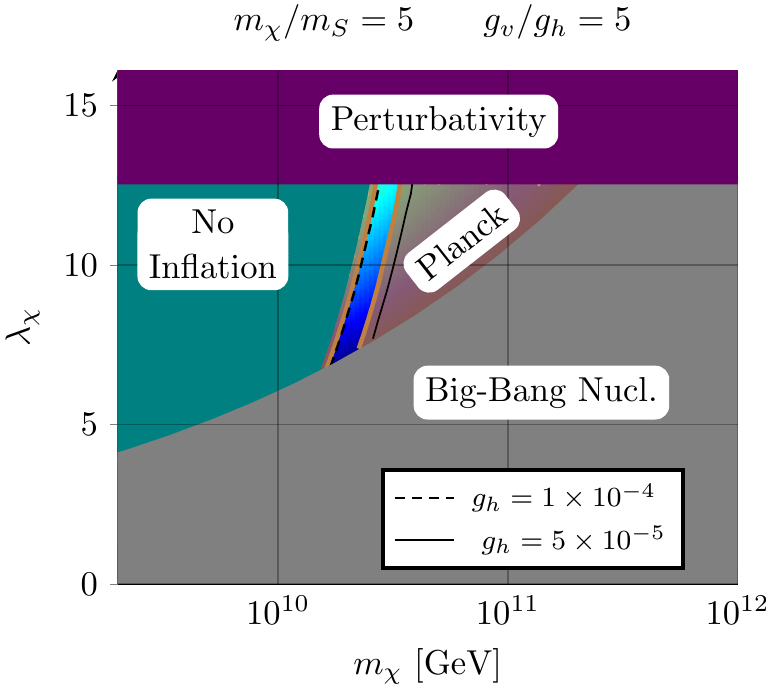}
\end{subfigure}
\begin{subfigure}[b]{0.437\textwidth}
        \centering
        \includegraphics[width=\linewidth]{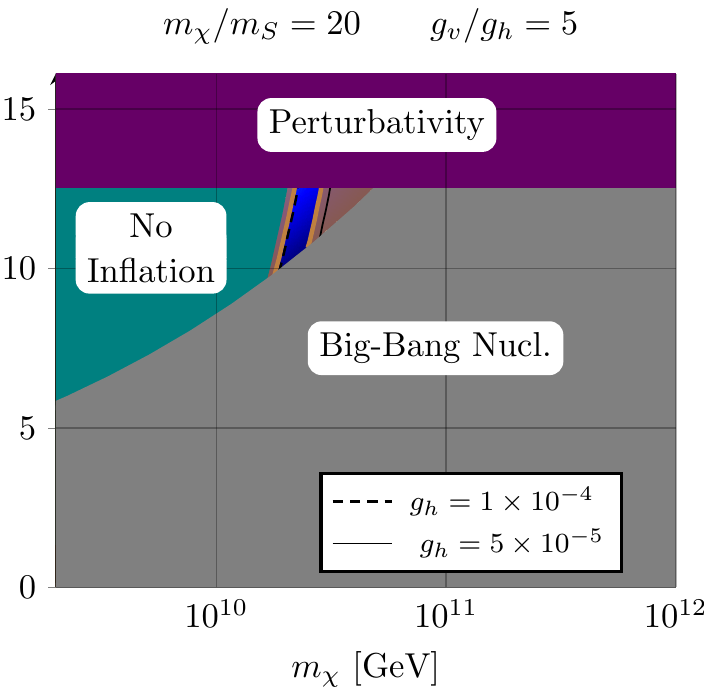}
\end{subfigure}\\
\begin{subfigure}[b]{0.4\textwidth}
        \centering
        \includegraphics[width=\linewidth]{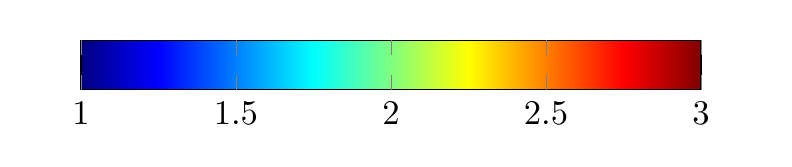}
\end{subfigure}\\
\vspace{-0.3cm}
$\mathrm{Log}_{10}[T_{\text{RH}}/\mathrm{MeV}]$
\caption{\label{fig:money}Fully constrained parameter space in which {each point coloured by the rainbow colour scheme} is giving the correct relic abundance of dark matter, satisfies the BBN limit and can {accommodate} a sufficient amount of e-folds for inflation to be successful.}
\end{figure*}

{\section{Dark-Matter Detection}
\label{sec:detection}
As was already noticed in Refs.~\cite{hooper1, hooper2}, such model of highly decoupled dark-matter, by construction, does not have any interesting signature for direct detection since the assumption that the SM and the dark sector are never in thermal equilibrium {implies that the scattering cross-section between the dark- and visible-sector particles are extremely small}. 
From the point of view of dark-matter indirect detection, {such models could potentially lead to interesting indirect detection signals due to the annihilaton of DM particles \cite{Cirelli:2018iax}.
However, in our scenario, the large mass ($m_{\chi}\gtrsim 10~\mathrm{EeV}$) for our DM particles significantly reduces their density in the galaxy and renders the indirect detection of any annihilation product impossible for any next generation detection experiment.}

{However one should note that the possibility that dark-matter particles decay directly into SM particles is not forbidden in our scenario as long as its lifetime is {sufficiently} larger than the age of the universe, and as long as the decay width of the dark scalar $S$ is not significantly affected by such a decay channel for DM particles. {Given the very large mass of DM particles considered here, imposing a dark-matter lifetime larger than the age of the universe would require any decaying process to involve extremely small couplings. Therefore such decay may not interfere with the entropy dilution process, and the decay of a DM candidate as heavy as $\sim 10~\mathrm{EeV}$ could be potentially detected through Ultra-High-Energy Cosmic Rays searches (UHECR).} Interestingly, such possibility has already been considered in Refs.~\cite{Heurtier:2019git, Hooper:2019ytr} where the decay of a dark-matter particle of mass  $\gtrsim \mathrm{EeV}$ into long-lived boosted particles interacting in the Earth with the electro-weak sector could explain anomalous up-going events seen by both the IceCube and ANITA collaborations~\cite{Fox:2018syq, Romero-Wolf:2017ope}. Such considerations are model dependent and are not explored in this paper. However, our scenario provides a dynamical way to generate such a heavy dark-matter particle population in the early universe.}

{
Another potential source of observational signatures comes from the structure formation during the EMD epoch.
In the standard cosmological history, after the inflaton decays, the universe is dominated by radiation until the recent matter-radiation-dominated (MD) epoch at $z~\sim~3400$.
In the radiation-dominated epoch (RD), the primordial density perturbation modes which have already entered the horizon cannot grow significantly, whereas their growth is rapid during the MD epoch.
Therefore, within the standard $\Lambda$CDM cosmological framework, the structure formation only takes place in the MD epoch.
However, such a situation can be significantly altered if an EMD epoch takes place after the reheating and before BBN.
In the EMD epoch, perturbation modes that have entered the horizon can grow and form structures.
One possibility is that,  as compared with the standard cosmology, a larger amount of primordial black holes (PBH) can be produced in the EMD era \cite{Georg:2016yxa,Georg:2017mqk, Georg:2019jld, Carr:2017edp, Carr:2018nkm}.
The PBHs are potentially detectable because they can evaporate and produce photons due to Hawking radiation \cite{Polnarev:1986bi}.
The bounds on such signatures have already been thoroughly studied, and severely constrain the abundance of PBH over a larger range of mass scales \cite{Carr:2009jm,Clark:2016nst}.
However, to have a signature that is large enough to be reached by present or near-future observations, even with the EMD epoch, the inflationary sector must be able to produce a blue-tilted primordial spectrum \cite{Georg:2016yxa,Georg:2017mqk, Carr:2017edp}.
This is not the case with the inflation models that we have chosen in this paper.
Nevertheless, a different inflation model could be chosen to couple to the dark and visible sectors, in which case the bounds on PBH could be applied.
The other possibility is to have an increased abundance of compact microhalos due to the rapid growth of density perturbation in the EMD epoch.
Since those microhalos could significantly boost the DM annihilation rate \cite{Erickcek:2011us,Barenboim:2013gya,Erickcek:2015jza,Erickcek:2015bda,Erickcek:2017zqj,Dror:2017gjq}, they could potentially compensate the fact that our visible and invisible sectors are highly decoupled and provide important indirect detection signatures.
However, the study of those indirect-detection signals requires a specific formulation of our Lagrangian in the visible sector.
Therefore, possible signatures from both PBHs and compact microhalos are very model-dependent, and a systematic study of these signatures is beyond the scope of this paper.}

\section{Conclusions and Comments}\label{sec:Conclusion}

In this paper we have explored the possibility that the inflaton may serve as the mediator through which the dark and visible sectors communicate. In the context of large-field inflation models, the fact that the mass scale required during inflation to match observations is about $10^{13}$ GeV suggests that no thermal equilibrium can ever take place between the dark and the visible sectors. Although  ``inflaton-portal" scenarios have been proposed in the literature to account for a freeze-in production mechanism of dark matter \cite{dev, moi}, it was shown that the freeze-in production of dark matter through the ``inflaton portal" has to be extremely fine-tuned in order to make it dominate over a direct decay of the inflaton into dark-matter. We have gone one step further in this work and considered the inflaton-portal scenario in models where, even though the dark sector is highly decoupled from the SM, one of its components eventually decays at very late time into SM particles and dilutes the dark-matter relic abundance. We have shown that introducing explicitly the inflaton as the only particle in contact with both thermal baths provides a way to obtain the two basic ingredients that such scenarios rely on in one stroke: $(i)$ The way the inflaton couples to the dark sector and the visible sector allows one to calculate the amount of energy transfered to both baths at the time of reheating. $(ii)$ The inflaton-portal interactions provide a decay channel for the lightest dark particle into SM particles which is naturally suppressed by the inflation propagator. As we have studied in detail, the fact that both this small decay rate and the temperature ratio of the two baths are intimately related in our scenario leads to non-trivial constraints on the parameter space. In particular, we have demonstrated that the correct dark-matter relic abundance can be obtained for an inflaton of mass \mbox{$10^{13}$ GeV} and with couplings of $\mathcal{O}(1)$ to both the visible and the dark sectors. The mass of the dark-matter particle is required in this case to be larger than $\mathcal{O}(10)~\mathrm{PeV}$ to {accommodate} all the low energy constraints of the model.

{From the point of view of inflation, we have also studied the radiative corrections to the inflation potential that the coupling of the inflaton to the matter sector may generate. 
As an important feature of the model, the complete knowledge of the {cosmological} evolution from the time of inflation to the present time, and in particular the existence of an early period of matter domination before the entropy dilution mechanism takes places, affects the number of e-folds of inflation required between horizon crossing and the end of inflation. Therefore, we have showed that the tension between the stability of the inflation trajectory, the BBN bound and the 1$\sigma$ exclusion limit from Planck completely rule out the possibility of a chaotic inflation scenario, while the parameter space remains open in the case of plateau-like inflation models such as $\alpha$-attractor models. However, we have seen that the Planck constraints already partially rule out part of the parameter space in this case and might be able to completely rule out our scenario in the future.
As a general feature of the model, we have seen that a relative proximity between the masses of the particles in the dark sector {$m_S\lesssim m_{\chi}$} and couplings of the inflaton between the dark and visible sector of order $g_v\lesssim g_h \sim \mathcal O(10^{-6}-10^{-4})$ is favored in our scan over the parameter space. Furthermore, it turns out that the dark-matter mass in our scenario is required to be roughly larger than $10~\mathrm{EeV}$ which might be consistent with {the} recent DM interpretation of the ANITA anomalous events~\cite{Heurtier:2019git}.
}

A few comments are in order. In this paper we chose to consider a class of $\alpha$-attractors, allowing us to consider both the case of a chaotic-inflation scenario ($\alpha\gg 1$), and the case of a plateau-inflation scenario ($\alpha\lesssim 1$). This has led us to rule out the case of chaotic inflation and favour plateau-inflation models for which we have showed that our dark-matter candidate must be heavier than $10$ EeV. A plethora of large-field inflation models exist and make predictions for observables such as the tensor-to-scalar ratio and the spectral index of the CMB spectrum which can also be in good agreement with recent measurements. The way radiative corrections may alter the inflation potential might lead to different constraints on the inflaton couplings to the other sectors and potentially relax or strengthen the aforementioned constraints on the dark-matter mass. Moreover, the mass of the inflaton in the vacuum at the end of inflation might be sensibly different in such cases {and might lead to potentially different constraints on the masses of the dark-sector particles.} In particular, lowering the inflaton mass can allow one to lower the dark-matter mass, which might render indirect-detection signatures of our model more promising in the near future. Therefore, a systematic review of different inflation scenarios would be necessary to explore all the possibilities that the inflaton-portal paradigm opens up. However, the reader should note that, in many of the models beyond the simplest inflationary scenarios, the mass of the inflaton in the vacuum can be chosen independently from the inflation energy scale. This is for instance the case in models involving a non-minimal coupling to gravity. This disconnection in scales may therefore render an inflaton-portal scenario poorly predictive. This is a typical problem that most of the studies attempting to connect inflation to low-energy phenomenology encounter. Nevertheless, one should note that most of the large-field inflation models involving only one mass scale in the inflationary sector predict a mass for the inflaton of order $10^{13}~\mathrm{GeV}$, {which is why we have chosen this benchmark value for the inflaton mass in our study}.

Moreover, the ratio $m_{\chi}/m_S$ has been taken throughout our results to be of order $\mathcal O( 1-10)$ in order not to introduce an arbitrary (and unnatural) hierarchy of masses in the dark sector. We have seen that such choice is favored by the different constraints we have considered. We point out that if the scalar $S$ would provide its mass to dark matter through a spontaneous symmetry breaking, a ratio of $\sim 1/10$ between $m_{\chi}$ and $m_S$ could be consistent with the $\mathcal O(1)$ Yukawa couplings $\lambda_s$ and $\lambda_p$ we have introduced. Such a possibility is outside the scope of this paper and would involve many subtleties. For example, if the hidden scalar couples to the Higgs or to the right-handed neutrino, providing $S$ a vacuum expectation value may have additional consequences for the phenomenology of our model.

{Finally, the presence of a long matter-domination era might lead the reader to worry about the plausibility of realizing leptogenesis/baryogenesis, when the Universe history would be significantly altered as compared to $\Lambda$CDM down to small temperature values. In our scenario, since the universe is dominated by the dark scalar, most of the visible bath is diluted until the dark scalar decays into SM particles and the universe is fully reheated again. It is after such a late reheating that we expect the baryon asymmetry to be generated in the Universe. As a matter of fact, this reheating temperature can be as large as $\mathcal{O}(10)~\mathrm{GeV}$ in our scenario, which allows for a low scale leptogenesis to take place, as is suggested in Refs.~\cite{Davoudiasl:2010am,Davoudiasl:2015jja,Aitken:2017wie,Dimopoulos:1987rk}.}

{We conclude by emphasizing that our paper opens up the possibility for a large class of models in which the inflaton portal can mediate interactions between a heavy dark-matter candidate and the visible sector. While such large dark-matter masses may be difficult to probe experimentally through direct or indirect detection in a near future through the inflaton portal, {we saw that observational bounds on the inflation sector derived by Planck from the study of the primordial-perturbation power spectrum are able to put constraints on our scenario.} 
The possibility that dark-matter particles decay directly into SM final states under the form of UHECR might finally provide a window to test the presence of such heavy dark matter in the universe.}

\vspace{20pt}
\noindent {\bf Acknowledgments. }  The authors would like to thank K. Dienes, E. Dudas, Y. Mambrini and I. Baldes for useful comments. The authors are extremely grateful to B. Thomas for his careful reading and extensive comments about the draft. L.H. would like to thank Doojin Kim, Luka Leskovec and Peter J.  Zimmerman for fruitful discussions. The research activities of LH and FH are supported  by the Department of Energy under Grant DE-FG02-13ER41976/DE-SC0009913.
\appendix
\section{Annihilation cross section}\label{AppendixA}
\noindent
The Lagrangian in Eq.~\eqref{eq:lagrangianHidden} gives rise to the t-channel annihilation cross section for the dark-matter particle $\chi$ as found in Ref.~\cite{Berlin:2015ymu}
\begin{eqnarray}
&&\sigma v(\chi\chi\to S S)=\frac{\lambda_s^2 \lambda_p^2 m_{\chi}^2}{2 \pi  \left(2 m_{\chi}^2-m_S^2\right)^2}\sqrt{1-\frac{m_S^2}{m_{\chi}^2}}\nonumber\\
&+&v^2 \frac{\sqrt{1-{m_S^2}/{m_{\chi}^2}}}{96 \pi  m_{\chi}^2 \left(2 m_{\chi}^2-m_S^2\right)^4} \mathlarger{\mathlarger{\mathlarger{\mathlarger{\left[\right.}}}}\left.2 m_{\chi}^2 \left(\lambda_p^4-\lambda_s^4\right) \left(m_S^2-m_{\chi}^2\right)^3\right.\nonumber\\
&+&3 m_{\chi}^2 \left(\lambda_s^2+\lambda_p^2\right)^2 \left(m_{\chi}^2-m_S^2\right)^3\nonumber\\
&+&\left. 6 m_{\chi}^6 \left(\lambda_p^4-\lambda_s^4\right) \left(m_S^2-m_{\chi}^2\right)\right.\nonumber\\
&-&\left.m_{\chi}^2 \left(m_S^2-m_{\chi}^2\right)^2 \left[m_{\chi}^2 \left(-5 \lambda_s^4+18 \lambda_s^2 \lambda_p^2-5 \lambda_p^4\right)\right.\right.\nonumber\\
&+& \left. 3 m_S^2 \left(\lambda_s^4-4 \lambda_s^2 \lambda_p^2+\lambda_p^4\right)\right]-8 m_{\chi}^4 \left(\lambda_p^4-\lambda_s^4\right) \left(m_S^2-m_{\chi}^2\right)^2\nonumber\\
&+&3 m_{\chi}^6 \left[m_{\chi}^2 \left(\lambda_p^2-\lambda_s^2\right)^2-m_S^2 \left(\lambda_s^4-4 \lambda_s^2 \lambda_p^2+\lambda_p^4\right)\right]\nonumber\\
&+&3 m_{\chi}^4 \left(m_S^2-m_{\chi}^2\right) \left[2 m_S^2 \left(\lambda_s^4-4 \lambda_s^2 \lambda_p^2+\lambda_p^4\right)\right.\nonumber\\
&-&\left.\left.3 m_{\chi}^2 \left(\lambda_s^4-6 \lambda_s^2 \lambda_p^2+\lambda_p^4\right)\right]\right.\mathlarger{\mathlarger{\mathlarger{\mathlarger{\left.\right]}}}}+\mathcal{O}(v^4)\,,
\end{eqnarray}
\section{\label{sec:RelicAbundance}Relic Abundance}
We can estimate the relic abundance analytically by making the following assumptions : 

\begin{itemize}
\item The dark matter decouples at $x_\text{FO}=\frac{m_{\chi}}{T_\text{FO}}$. Its number density is obtained from $Y_{\chi}(\infty)$.
\item The scalar $S$, in the absence of number depleting processes, also decouples at $x_\text{FO}$ while its phase-space distribution is still thermal. Therefore its comoving number density is given by $Y_S(x_\text{FO})=Y_S^{eq}(x_\text{FO})$.
\end{itemize}
First of all, let us approximate the value of the decoupling temperature. From Ref.~\cite{hooper2} one can write the value of $x_\text{FO}$  analytically for an arbitrary cross-section of the form $\langle\sigma v\rangle \approx a +  b\ v^2$ as
\begin{equation}
x_\text{FO}=\xi\ln \left[\frac{c(c+2)}{4\pi^3}\sqrt{\frac{45}{2}}\frac{g_{\chi}}{\sqrt{g_{\star}^\text{eff}}}m_{\chi}m_{p}\frac{\xi^{5/2}(a+\frac{6\xi b}{x_\text{FO}})}{\sqrt{x_\text{FO}}(1-3\frac{\xi}{2x_\text{FO}})}\right]\,,
\end{equation}
where we take $c\approx 0.4$ \cite{Kolb:1990vq}. Expanding this expression by assuming $\xi/x_\text{FO}\ll 1$, which is a good approximation in the parameter space of interest, one finds that
\begin{equation}\label{eq:B2}
x_\text{FO}\approx\xi\left[38.8+\ln\left(\frac{a g_{\chi}m_{\chi}}{\mathrm{GeV^{-1}}}\right)+\ln\left(\frac{\xi^2}{\sqrt{g_{ \star}^\text{eff}}}\right)\right]\,.
\end{equation}
Assuming $a\lesssim m_{\chi}^{-2}$, the second term in the square brackets in Eq.~\eqref{eq:B2} can be safely neglected.
Solving the Boltzmann equation for $\chi$ gives \cite{hooper2}
\begin{equation}\label{eq:B3}
Y_{\chi}^{-1}=\sqrt{\frac{\pi}{45}}g_{\star}m_{\chi}m_p\int_{x_\text{FO}}^{\infty}dx \frac{a+6\xi b/x}{x^2\sqrt{g_{\star}^\text{eff}}}\,,
\end{equation}
where $g_{\star}^\text{eff}=g_{\star}+\xi^4 g_{\star}^h$. At freeze-out, the value of the temperature ratio is given by \cite{hooper2} 
\begin{eqnarray}
\xi_\text{FO}&\approx&\left(\frac{g_{\star,\text{inf}}^{h}}{g_{\star}^h(T^h_\text{FO})}\right)^{1/3}\xi_{\text{inf}}=\left(1+\frac{c_{\chi}g_{\chi}}{c_S g_S}\right)^{1/3}\xi_{\text{inf}}\nonumber\\
&=&\left(1+\frac{c_{\chi}g_{\chi}}{c_S g_S}\right)^{1/3}\left(\frac{g^{\star}_{\text{ inf}}}{g^{\star}_{h,\text{ inf}}}\right)^{1/4}\times\left(\frac{g_{h}}{g_{v}}\right)^{1/2}\,.\nonumber\\
\end{eqnarray}
Thereafter the temperatures ratio has been shown to decrease in proportion to the temperature \cite{hooper2} such that
 \begin{equation}\label{eq:xievolution}
\xi=\xi_\text{FO}\left(\frac{x_\text{FO}}{x}\right)=\left(1+\frac{c_{\chi}g_{\chi}}{c_S g_S}\right)^{1/3}\xi_{\text{inf}}\left(\frac{x_\text{FO}}{x}\right)\,.
\end{equation}
In evaluating Eq.~\eqref{eq:xievolution}, there are two regimes of interest, which correspond to different relationships between the quantity $g_{\star}^h \xi^4$ and $g_{\star}$.  We consider these regimes as two separate cases below.

\vspace{5pt}
\paragraph*{$\bullet$ $g_{\star}^h\xi^4\ll g_{\star} $} ~\vspace{5pt}\\ In this case $g_{\star}^\text{eff}$ is dominated by $g_{\star}$. Using \eqref{eq:xievolution} one finds\footnote{Note the difference of factor in front of the term in $b$ as compared to \cite{hooper2} due to the fact that this reference use a constant $\xi$ after freeze-out of dark matter whereas we use an inverse power law due to the simultaneous decoupling of $S$ in our case.}
\begin{equation}\label{eq:relicchi}
Y_{\chi}^{-1}=\sqrt{\frac{\pi}{45}}g_{\star}m_{\chi}m_p \frac{a+2\xi_\text{FO} b/x_\text{FO}}{x_\text{FO}\sqrt{g_{\star}}}\,.
\end{equation}
In the case in which $b\lesssim a$, the relic density before entropy dilution can be written as
\begin{equation}
\Omega_{\chi}h^2 = \frac{m_{\chi}s_0 Y_{\chi}}{\rho_c}\approx 8.5\times 10^{-11}\frac{x_\text{FO}}{\sqrt{g_{\star}}}\frac{\mathrm{GeV}^{-2}}{a}\,.
\end{equation}
According to Ref.~\cite{Kolb:1990vq}, the entropy-dilution factor is computed to be
\begin{equation}
S_f/S_i=1.83 g_{\star}^{1/4}\frac{m_S Y_S(x_f)\tau_S^{1/2}}{m_p^{1/2}}\,.
\end{equation}
Defining the reheating temperature resulting from the decay of $S$ as
\begin{equation}
T_{\text{RH}}\equiv\left(\frac{90}{8\pi^3 g_{ \star}}\right)^{1/4}\sqrt{\Gamma_S m_p}\,,
\end{equation}
one finds that
\begin{equation}
S_f/S_i=1.83 \left(\frac{90}{8\pi^3}\right)^{1/4}\frac{m_S Y_S(x_\text{FO})}{T_\text{R}}\,.
\end{equation}
As mentioned above, the yield of $S$ at the reheating time can be computed using the approximate equilibrium number density
\begin{equation}
Y_S(x_\text{FO})=\frac{g_S}{2\pi^2}\frac{m_S^2 \xi T_\text{FO} K_2(m_S/\xi T_\text{FO})}{s(T_\text{FO})}\,.
\end{equation}
Using the fact that the argument of the modified Bessel function is small in our parameter-space regime of interest\footnote{Note that for a ratio $m_{\chi}/m_S$ going roughly going under $\sim 21.3$ the argument of the Bessel function  becomes bigger than one and the numerical error may become important.}, one can approximate $K_2(x\ll 1)\approx 2/x^2$. Making this approximation and noting that $s(T) = 2\pi^2 g_{\star}^s T^3/45$, we get
\begin{equation}
Y_S(x_\text{FO})\approx\frac{45}{2\pi^4}\frac{g_S}{g_{\star}^s}\xi^3\approx 2.18\times 10^{-3}\xi^3\,.
\end{equation}
Putting everything together and using $\xi=\xi_\text{FO}$ at freeze-out, we arrive at our final result:
\begin{eqnarray}\label{eq:analytic}
\frac{\Omega_{\chi}h^2}{S_f/S_i}&\approx& 0.16\left[1+0.04\ln\left(\frac{\xi_\text{FO}^2}{\sqrt{g_{\star}+\xi_\text{FO}^4 g_{\star}^h}}\frac{\alpha_{\chi}^2}{0.8^2}\frac{50\mathrm{PeV}}{m_{\chi}}\right)\right]\nonumber \\
&\times&\xi_\text{FO}^{-2} \left(\frac{T_\text{R}}{10\mathrm{MeV}}\right)\left(\frac{m_{\chi}}{50\mathrm{PeV}}\right)\left(\frac{m_{\chi}/m_S}{20}\right)\left(\frac{0.8}{\alpha_{\chi}}\right)^2\,.\nonumber\\
\end{eqnarray}
\paragraph*{$\bullet$ $g_{\star}^h\xi^4\gg g_{\star} $ }
~\vspace{10pt}\\
The denominator of the integral in Eq.~\eqref{eq:B3} varies significantly between $x_\text{FO}$ and the point $\tilde{x}\equiv x_\text{FO}(g_h/g_v)^{1/2}$ where $g_{\star}^h\xi^4\approx g_{\star}$. We therefore separate the integral into two parts as follows
\begin{eqnarray}
\int_{x_\text{FO}}^{\infty}dx \frac{a+6\xi b/x}{x^2\sqrt{g_{\star}^\text{eff}}}&=&\int_{x_\text{FO}}^{\tilde{x}}dx \frac{a+6\xi b/x}{x^2\sqrt{\xi^4 g_{\star}^{h}}}\nonumber\\
&+&\int_{\tilde{x}}^{\infty}dx \frac{a+6\xi b/x}{x^2\sqrt{g_{\star}}}\,,
\end{eqnarray}
where $\xi$ still evolves as in Eq.~\eqref{eq:xievolution}.  This integral can be calculated to be
\begin{eqnarray}
\int_{x_\text{FO}}^{\infty}dx \frac{a+6\xi b/x}{x^2\sqrt{g_{\star}^\text{eff}}}&=&\frac{a}{\tilde{x}\sqrt{g_{\star}}}\left(2-\frac{x_\text{FO}}{\tilde{x}}\right)\nonumber\\
&+&\frac{6 \xi_\text{FO}b}{\tilde{x}^2\sqrt{g_{\star}}}\left(1-\frac{2}{3}\frac{x_\text{FO}}{\tilde{x}}\right)\,.
\end{eqnarray}
From this formula, we see that taking the limit where $\tilde{x}=x_\text{FO}$ brings us back to the the expression in Eq.~\eqref{eq:relicchi}. Again one can safely neglect the $b$-term as long as $b\lesssim a$ and $\xi_\text{FO}/\tilde{x}\ll 1$, which are reasonable assumptions since $\xi_\text{FO}/x_\text{FO}\approx 21$ and $\tilde{x}>x_\text{FO}$.

Putting everything together, we get 
\begin{equation}
Y_{\chi}^{-1}=\sqrt{\frac{\pi}{45}}g_{\star}m_{\chi}m_p \frac{a}{\tilde{x}\sqrt{g_{\star}}}\left(2-\frac{x_\text{FO}}{\tilde{x}}\right)\,,
\end{equation}
which, in comparison with the previous case, represents a rescaling of the relic density by a factor
\begin{equation}
\frac{(Y_{\chi}^{-1})_{g_{\star}^h\xi^4\gg g_{\star}}}{(Y_{\chi}^{-1})_{g_{\star}^h\xi^4\ll g_{\star}}}=\frac{x_\text{FO}}{\tilde{x}}\left(2-\frac{x_\text{FO}}{\tilde{x}}\right)=\sqrt{\frac{g_v}{g_h}}\left(2-\sqrt{\frac{g_v}{g_h}}\right)\,.
\end{equation}
The final relic density in this case is simply
\begin{eqnarray}
&&\frac{\Omega_{\chi}h^2}{S_f/S_i}\approx \text{Eq.~\eqref{eq:analytic}}\times \sqrt{\frac{g_v}{g_h}}\left(2-\sqrt{\frac{g_v}{g_h}}\right)\,.\nonumber\\
\end{eqnarray}

{\section{\label{sec:efolds}Number of e-folds and Early Matter Domination}
Let us list in this appendix the expressions for the different energy densities which have been used in Eq.~\eqref{eq:efolds} in order to compute the number of e-folds. While inflation takes place, the slow roll conditions
\begin{equation}
\epsilon_V=\frac{M_p^2}{2}\left(\frac{V_\text{inf}'}{V_\text{inf}}\right)^2\ll 1\,,\quad\text{and}\quad \eta_V=M_p^2\left(\frac{V_\text{inf}''}{V_\text{inf}}\right)\ll 1\,,
\end{equation}
are satisfied and inflation ends when {$\epsilon_V(\phi_\text{end})=1$}. This defines {the inflaton field value} at the end of inflation where one can compute the energy density
\begin{equation}
\rho_\text{end}\approx V_\text{inf}(\phi_\text{end})\,.
\end{equation}
After the inflaton decay takes place the energy density of the universe is shared between the visible and hidden thermal baths. The energy density in the hidden sector is
\begin{equation}
\rho^h_{\text{reh}}=(c_{\chi}g_{\chi}+c_S g_S)\frac{\pi^2}{30}(T_\text{inf}^h)^4\,,
\end{equation}
while
\begin{equation}
\rho^v_{\text{reh}}=g_\star^{SM}\frac{\pi^2}{30}(T_\text{inf})^4\,,
\end{equation}
where $c_{\chi}=7/8$, $c_s=1$, $g_{\chi}=4$ and $g_S=1$. The hidden sector temperature is given by $T^h=\xi(T)T$ at any time and the dark sector energy density is proportional to $(T^h)^4$ when dark matter {freezes out} and until $S$ becomes non-relativistic. Right after freeze-out, entropy conservation provides
\begin{equation}
\xi_\text{FO}=\left(1+\frac{c_{\chi}g_{\chi}}{c_Sg_S}\right)^{1/3}\xi_\text{inf}\,,
\end{equation}
and the hidden energy density is therefore given by
\begin{eqnarray}
\rho^h_\text{FO}&=&c_Sg_S \frac{\pi^2}{30}(\xi_\text{FO} T_\text{FO})^4=\rho^h_\text{reh}\left(\frac{T_\text{FO}}{T_\text{inf}}\right)^4\left(1+\frac{c_{\chi}g_{\chi}}{c_Sg_S}\right)^{1/3}\,,\nonumber\\
\end{eqnarray}
where we introduced the visible sector temperature at DM freeze-out $T_\text{FO}$ as defined in Appendix B. The energy density for the dark sector after dark matter freezes out evolves like $(T^h)^4\propto T^4$ before $S$ becomes {non-relativistic} around $T_S\equiv m_S/\xi_\text{FO}$, after which the energy density scales like $\propto T^3$, giving for $T<T_S$ 
\begin{equation}
\rho^h_S(T<T_S)=\rho_\text{FO}^h\times\left(\frac{T_S}{T_\text{FO}}\right)^4\left(\frac{T}{T_S}\right)^3\,.
\end{equation}
Assuming that matter domination {takes} place early enough such that {the effective number of relativistic degrees of freedom does not change}, the visible bath energy density simply evolves as
\begin{equation}
\rho^v(T)=\rho^v_\text{reh}\times \left(\frac{T}{T_\text{inf}}\right)^4\,,
\end{equation}
and the early matter-domination period starts when the dark scalar {starts} dominating the energy density, which happens at
\begin{equation}
T_\text{EMD}=\frac{\rho_\text{FO}^h}{\rho_\text{inf}^v}\times\left(\frac{T_\text{inf}}{T_\text{FO}}\right)^4 \frac{m_S}{\xi_\text{FO}}\,.
\end{equation}
The energy density at this moment is given by
\begin{equation}
\rho_\text{EMD}\equiv 2 \rho^v(T_\text{EMD}) =2 \rho_S^h(T_\text{EMD})\,.
\end{equation}}

{Later on, the dark scalar decays and the universe is reheated with energy density
\begin{equation}
\rho_\text{dec}=g^\star_{SM} \frac{\pi^2}{30}T_\text{R}^4\,,
\end{equation}
where $T_\text{R}=(90/8\pi^3)^{1/4}\sqrt{\Gamma_S m_p}$ was defined earlier.}

\vfill
\pagebreak

\vfill
\end{document}